\documentclass[openacc]{rstransa}%%%%where rstrans is the template name
\usepackage{graphicx}

\begin{document}

%%%% Article title to be placed here
\title{Quantum Annealing: An Overview}

\author{%%%% Author details
Atanu Rajak$^{1}$, Sei Suzuki$^{2}$, Amit Dutta$^{3}$ and Bikas K. Chakrabarti$^{4,5}$}

%%%%%%%%% Insert author address here
\address{$^{1}$Presidency University, Kolkata -700073, India\\
$^{2}$ Saitama Medical University, Moroyama, Saitama 350-0495, Japan\\
$^{3}$Indian Institute of Technology Kanpur, 208016, India\\
$^{4}$ Saha Institute of Nuclear Physics, 1/AF Bidhannagar, Kolkata-700064, India\\
$^{5}$ Indian Statistical Institute, 203   B. T. Road, Kolkata-700108, India}

%%%% Subject entries to be placed here %%%%
\subject{quantum annealing and optimisation, quantum phase transitions, disordered systems, open quantum systems }

%%%% Keyword entries to be placed here %%%%

%%%% Insert corresponding author and its email address}
\corres{Atanu Rajak\\
\email{atanu.physics@presiuniv.ac.in}}

%%%% Abstract text to be placed here %%%%%%%%%%%%
\begin{abstract}
In this review, after providing  the basic physical concept behind quantum annealing (or adiabatic quantum computation), we present an overview of  some recent theoretical
as well as experimental developments pointing to the issues which are still debated.  With a brief discussion on the  fundamental  ideas of continuous and discontinuous quantum phase transitions, we discuss the Kibble-Zurek scaling of  defect generation
following a ramping of a quantum many body system across a quantum critical point. In the process, we discuss associated  models, both pure and disordered, and shed light on implementations
and some recent applications of the quantum annealing protocols. Furthermore, we discuss the effect of environmental coupling on quantum annealing. Some  possible ways to speed up the annealing protocol in closed systems are elaborated
upon:
We especially  focus on the recipes to avoid discontinuous quantum phase transitions occurring  in some models where  energy gaps vanish exponentially with the system size. \end{abstract}
%%%%%%%%%%%%%%%%%%%%%%%%%%%

%%%%%%%%%% Insert the texts which can accomdate on firstpage in the tag "fmtext" %%%%%

%\begin{fmtext}
%\end{fmtext}

%%%%%%%%%%%%%%% End of first page %%%%%%%%%%%%%%%%%%%%%

\newcommand{\tr}{\textcolor{red}}
\newcommand{\tb}{\textcolor{blue}}

\maketitle
\section{Introduction}\label{sec:Introduction}
%%%% Insert A head here

Following the recent technological advance in manipulation of a quantum state, 
the notion {of} quantum computation and simulation which initially stemmed from pure theoretical concepts has now {spread to flourish an} 
industry with an immense possibility of technological applications. In particular, studies of quantum annealing (QA) have gained a tremendous momentum
since programmable QA machines, dubbed as quantum annealers,
with more than thousands of qubits have been realized and commercialized.
In this review, having  provided an overview of QA protocol, we  discuss some recent theoretical and experimental developments of the QA
exploiting the advantage of utilizing quantum tunneling in finding the minimum of a classical energy function.

QA is usually aimed at seeking the ground state of a generic Ising model, which may contain random biases and/or random many-body interactions \cite{finnila_quantum_1994,kadowaki_quantum_1998,farhi_quantum_2001}. Many optimization problems including the traveling salesman problem, 
job scheduling problem, knapsack problem, and so on,
are shown to reduce to this problem.
Therefore the application of QA extends from physics to our daily life. This broadness of application is another reason why QA is attracted much attention in industry.
Now, let us consider an Ising model denoted by the Hamiltonian $H_P$, where the subscript $P$ stands for the problem Hamiltonian. We assume that $H_P$ is a classical many-body Ising Hamiltonian described in terms of the $z$ components of the Pauli operator $\{\sigma_j^z\}$. We further introduce a driver Hamiltonian $H_D$ which is not commutative with $H_P$ and has the trivial ground state. A simple choice for $H_D$ is the transverse field: 
$H_D = -\sum_j\sigma_j^x$, so that $H_D$ does not commute with $H_P$. The total Hamiltonian of QA is given as
\begin{equation}
 H(t) = A(t)H_D + B(t)H_P,
\label{eq:QA_Hamiltonian}
\end{equation}
where $A(t)$ and $B(t)$ are the scheduling function satisfying $A(t_i)\gg B(t_i)$ at the initial time $t_i$ and $A(t_f)\ll B(t_f)$ at the final time $t_f$ so that $H(t)$ interpolates between $H_D$ at $t=t_i$ and $H_P$ at $t=t_f$. The initial state at $t=t_i$ is set at the ground state of $H_D\approx H(t_i)/A(t_i)$. If the change in $H(t)$ with $t$ is ``sufficiently'' small, the spin state evolves adiabatically (i.e., stays in the ground state of the instantaneous Hamiltonian) and arrives at the ground state of $H_P$ at $t=t_f$ which we seek. This constitutes the basic notion of the  QA,  also known as the adiabatic quantum computation \cite{santoro_optimization_2006,das_colloquium_2008,morita_mathematical_2008,tanaka_quantum_2017,albash2018adiabatic,hauke2020perspectives,Das_2005}. 
Throughout this paper, we shall employ  QA  scheme using the transverse Ising Hamiltonian (if not otherwise mentioned).  To illustrate, we consider the following Hamiltonian with ferromagnetic nearest neighbour interactions in one dimension:
\begin{equation}
 H = - J\sum_{j}\sigma_j^z\sigma_{j+1}^z - \Gamma\sum_j\sigma_j^x .
\label{eq:pure1dTIM}
\end{equation}
where $J$ denotes the strength of the interaction and $\Gamma$ is the strength of the  non-commuting transverse field. Here $H_D=-\sum_j\sigma_j^x$ and $H_P=-\sum_j\sigma_j^z\sigma_{j+1}^z$. The transverse field $\Gamma$ is annealed to reach the ground state of $H_P$ from the ground state of $H_D$.

The success of QA is determined by how slowly the Hamiltonian changes with time. According to the adiabatic theorem of quantum mechanics, the criterion of the adiabatic time evolution is given by \cite{messiah_quantum_1961}
\begin{equation}
 \frac{\max\left[|\langle 1(t)|\frac{dH(t)}{dt}|g(t)\rangle|\right]}
  {\min[\Delta(t)]^2} \ll 1,
\label{eq:AdiabaticThm}
\end{equation}
where $|g(t)\rangle$ and $|1(t)\rangle$ are the instantaneous ground and first-excited states at time $t$, respectively, and $\Delta(t)$ denotes the instantaneous energy gap above $|g(t)\rangle$. 
The $\min$ and $\max$ functions are taken with respect to the variable $t$.
Thus, roughly speaking, QA works better for larger $\Delta (t)$ \cite{suzuki_residual_2005}.

As a classical counterpart to QA, simulated annealing (SA) is a known method of computation for optimization problems \cite{kirkpatrick_optimization_1983}. In this method, we prepare the Gibbs-Boltzmann distribution of $H_P$ at sufficiently high temperature by means of the Monte-Carlo method and literally anneal the system down to zero temperature. If annealing is sufficiently slow, then we are expected to arrive at the ground state of $H_P$ with high probability.
SA utilizes the thermal fluctuation for optimization, which induces the thermal (Arhenius) jump from a local energy minimum to another separated by an energy barrier. 
\begin{figure}[t]
\centering
%\begin{center}
  \includegraphics[width=5cm, bb = 0 0 268 281]{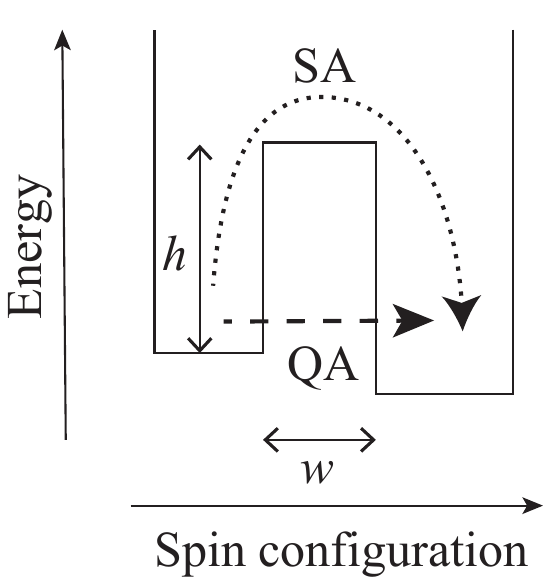}
% \end{center}
\caption{Schematic picture of the thermal fluctuation and quantum tunneling
in a system with local energy minima separated by an energy barrier with 
the height $h$ and the width $w$.}
\label{fig:barrier}
\end{figure}
The escape rate from a local minimum over the energy barrier with height $h$ is
given by $e^{-h/k_BT}$, where $k_B$ and $T$ denote the Boltzmann constant and
the temperature. Assuming that $h$ is proportional to the system size $N$, 
this suggests that an exponentially long time in $N$ is necessary to reach the global energy minimum by SA.

\begin{figure}
\begin{center}
 \includegraphics[width=13.5cm, bb=0 0 704 490]{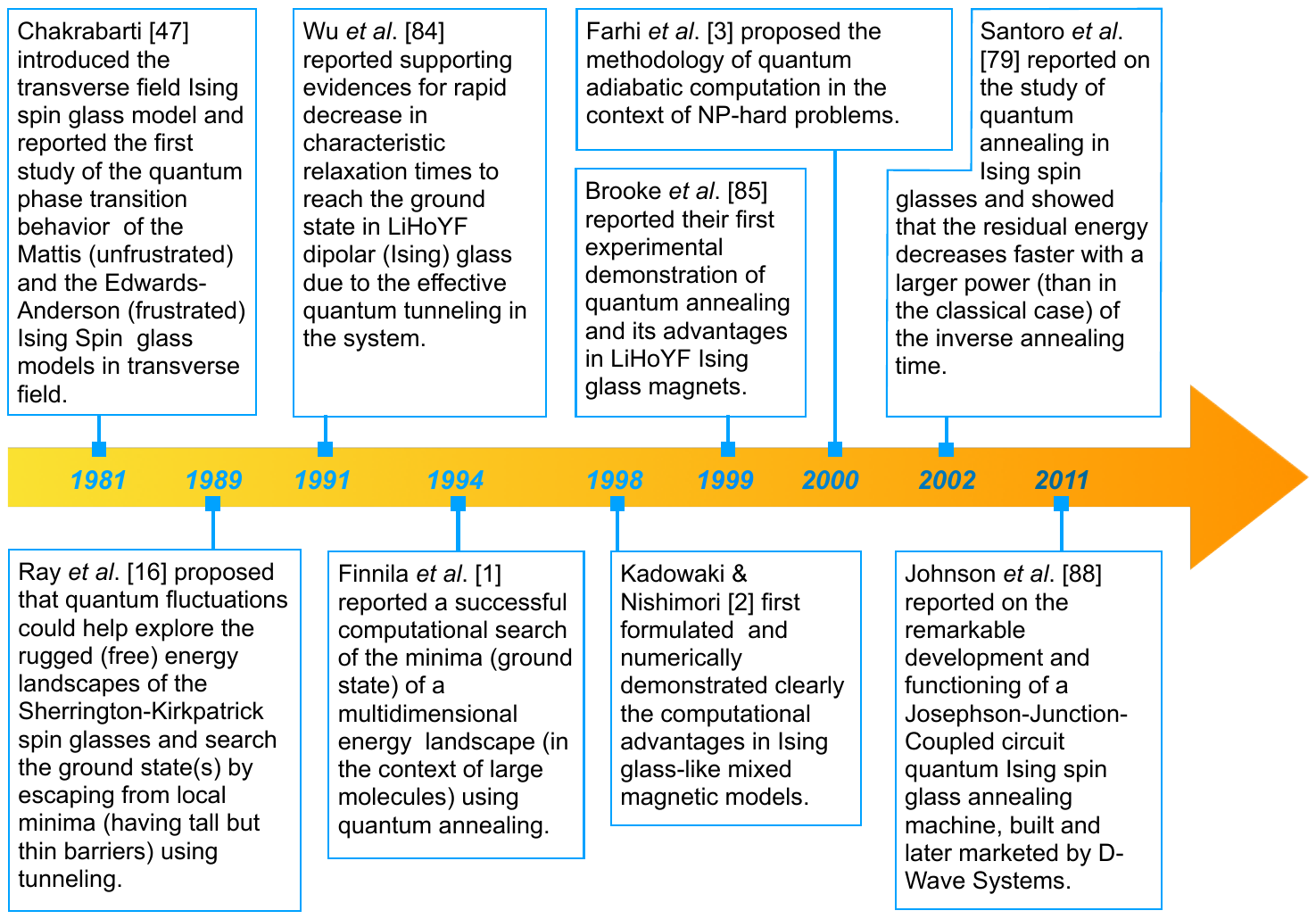}
\end{center}
 \caption{A brief Time-line for the development of Quantum Annealing.}
\label{fig:timeline}
\end{figure}

In contrast to SA, quantum tunneling induces an escape from a local minimum through
an energy barrier as shown schematically in Fig.~\ref{fig:barrier}. The tunneling probability
is approximately given by $e^{-\sqrt{h} w/g}$ \cite{shankar_principles_2012,mukherjee_multivariable_2015}, where $g$ denotes the strength of 
quantum fluctuation which corresponds to the transverse field $\Gamma$ in 
transverse Ising models. 
%We note that the tunneling probability 
%provides the minimum energy gap when two energy levels separated by a barrier degenerate. 
Therefore, assuming the height $h\sim O(N)$ and {the width} $w< O(N^{1/2})$, 
the time necessary to escape from a local minimum due to quantum tunneling is subexponential in $N$. For such a system, quantum tunneling helps the system to equilibrate even though the system is glassy, i.e., non-ergodic in the absence of the quantum fluctuation, leading to a potential advantage of QA over SA in glassy systems.
This role of quantum tunneling was first discussed by Ray et al., in 1989 \cite{Ray_1989}  (see discussions in  \cite{starchl_unraveling_2022}) in this regard)  in the context of the restoration of the replica symmetry or ergodicity due to quantum fluctuation in the quantum version of the Sherrington-Kirkpatrick model \cite{sherrington_solvable_1975}, which is detailed in the next section. 
Although  the existence of an energy landscape with thin and high barriers
in specific models is still an issue of debate, 
it must be a foundation for the speedup of QA over SA \cite{starchl_unraveling_2022,yaacoby_comparison_2022}.
In addition, several numerical and experimental studies have provided evidences for such an 
advantage of QA over SA in some specific models as shown in Secs. \ref{sec:Introduction}\ref{sec:TIM}\ref{sec:1dTIM} and \ref{sec:implementation}. We show a brief time-line for the development of QA in Fig. \ref{fig:timeline}.

The review is organized in the following fashion: Having discussed the basic idea behind the QA scheme and the results for various models in the context of annealing and defect
generation especially  for annealing across a  quantum critical point in Sec. \ref{sec:Introduction}, we move to discuss various  implementations of annealing protocols in Sec. \ref{sec:implementation}. In Sec. \ref{Sec:environment}. , we probe how does coupling to an external environment influence the QA process. In Sec. \ref{sp_qa}, we again  refer
to the close systems and discuss possible ways to speed QA processes especially in the context of avoiding discontinuous phase transitions. Some recent applications are
discussed in Sec. \ref{applications}.

\subsection{Quantum phase transition and quantum annealing}\label{sec:QPT_QA}
The minimum gap $\min[\Delta(t)]$ appearing in Eq.~(\ref{eq:AdiabaticThm}) often decreases with increasing the number of spins. In general the energy gap vanishes at a quantum phase transition (QPT) because a QPT separates disordered and ordered phases and the ground state is degenerate at a transition. Let us consider the transverse Ising Hamiltonian introduced in Eq.~(\ref{eq:pure1dTIM}). The initial state, for $\Gamma \gg J$, is given by the ground state of the transverse field, in which all spins are aligned along the $x$ axis of spin. This is a disordered state where {the ground-state} averaged  magnetization in the $z$ direction of spin is zero, i.e., $\langle \sigma_i^z\rangle =0$. The targeted ground state of the Ising Hamiltonian for $\Gamma=0$,  however, is an ordered state in the sense that it has a fixed magnetization $+1$ or $-1$ for each $\sigma_j^z$. This implies that the system encounters a QPT during QA. Indeed, the model in Eq.~\eqref{eq:pure1dTIM} has QPTs at $\Gamma/J =\pm1$. The finite  size scaling of the energy gap at QPT depends on the character of the associated QPT, and the latter is determined by the property of the Ising Hamiltonian.

The character of a conventional continuous QPT is specified by critical exponents \cite{sondhi1997continuous,sachdev_quantum_1999,suzuki_quantum_2012,dutta_quantum_2015}. The size scaling of the energy gap at a quantum critical point is given as $\Delta_c\sim L^{-z}$ where $L$ denotes the linear size of the system and the exponent $z$, known as the dynamical exponent, characterizes the associated  quantum critical point (QCP). Therefore the time for QA to work scales polynomially with the system size. However, apart from this simple situation, the polynomial scaling of the energy gap at QPT is not always true. In fact, a discontinuous QPT usually gives rise to an exponential scaling  with the system size. This can be understood phenomenologically as follows. Consider a quantum many-body system and focus on the two lowest energy levels. We assume that higher energy levels are highly separated from them.
The effective Hamiltonian is then written as
\[
 H = \left[
 \begin{array}{cc}
  \varepsilon_A& \mathit{\Delta}\\
 \mathit{\Delta} & \varepsilon_B\end{array}
\right] ,
\]
\begin{figure}[t]
 \begin{center}
  \includegraphics[width=8cm, bb = 0 0 468 418]{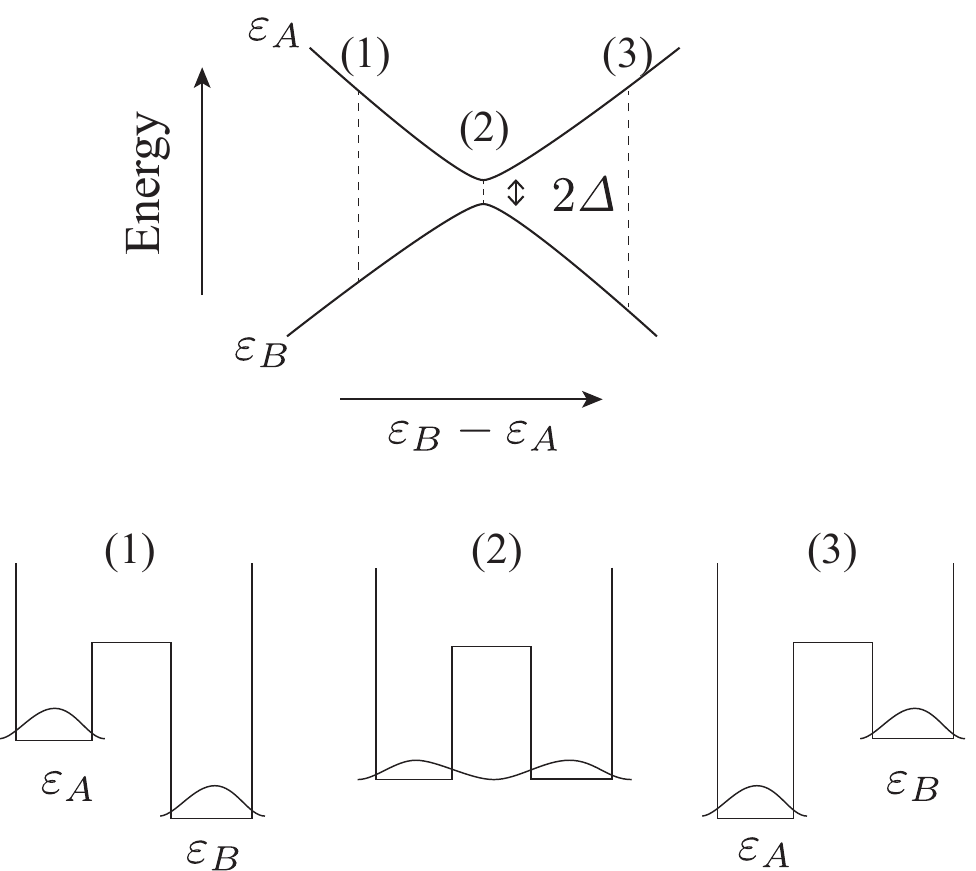}
 \end{center}
\caption{Schematic picture for an interchange of two energy levels.
$\varepsilon_A$ and $\varepsilon_B$ corresponds to the energies of the two local minima.
(1), (2), and (3) shows the situations with $\varepsilon_A\gg \varepsilon_B$, 
$\varepsilon_A=\varepsilon_B$, and $\varepsilon_A\ll\varepsilon_B$, respectively.
In the case (2) with $\varepsilon_A=\varepsilon_B$, the energy gap is given by
the twice of the tunneling energy $\mathit{\Delta}$ between the states $A$ and $B$.}
\label{fig:two-level-system}
\end{figure}
where $\varepsilon_A$ and $\varepsilon_B$ corresponds to the energies of the two local minima, and
$\mathcal{\Delta}$ denotes the tunneling energy between these two states. Figure \ref{fig:two-level-system} shows the energy levels of this Hamiltonian schematically.
The discontinuous QPT corresponds to the change of the lowest energy level between $A$ and $B$. The transition takes place where the bare energies $\varepsilon_A$ and $\varepsilon_B$ of two levels are degenerate. The energy gap at the transition is given by the twice of the tunneling energy $\mathit{\Delta}$, and $\mathit{\Delta}$ is given by an exponential of the Hamming distance between the states $A$ and $B$. Note that the Hamming distance is the number of sites at which the spin orientation along $z$ axis is different. Usually this distance increases linearly with the system size. Therefore the energy gap at a transition decays exponentially with the system size. Since the discontinuous QPT hinders QA, several ways to avoid the discontinuous QPT have been proposed. We will mention some of them in Sec. \ref{sp_qa}.

QA across a QPT is closely related to the Kibble-Zurek mechanism of defect generation following an annealing  across a QCP. \cite{kibble_topology_1976,zurek_cosmological_1985,zurek_dynamics_2005,polkovnikov_universal_2005,damski_simplest_2005,mukherjee_quenching_2007,sen_defect_2008}. The system starting from the initial disordered ground state evolves adiabatically as far as the characteristic time of the instantaneous ground state (i.e., the inverse of the gap) is shorter than  the annealing speed. However, on approaching a QPT, the characteristic time grows and hence the dynamics becomes
non-adiabatic in the vicinity of the QCP.
The state of the system after the passage through the QCP is no longer the ground state, rather a state with topological defects. The residual energy density, $\varepsilon_{\rm res}$, i.e., the excess energy over the expected final ground state at the end of the QA is a monotonically decreasing function of the annealing duration $\tau$. In case of a linear annealing through a conventional continuous QPT in a $d$-dimensional many-body system with the critical exponent $\nu$ for the correlation length and the dynamical exponent $z$, Kibble-Zurek scaling of the residual energy is given by
$\varepsilon_{\rm res}\sim \tau^{-d\nu/(z\nu+1)}$ {as far as the system after annealing is in a gapped phase}.  The scaling of the residual energy density is modified from the Kibble-Zurek scaling for other unconventional continuous QPTs or discontinuous QPTs and when the annealing protocol involves a non-linear variation of
the tuning parameter \cite{sen_defect_2008,barankov_optimal_2008,mukherjee_adiabatic_2010}. The scaling of the residual energy together with the scaling of the energy gap at QPT is an important measure that characterizes the property of QA \cite{dziarmaga_dynamics_2010,polkovnikov_colloquium_2011}.

\subsection{Transverse Ising models}\label{sec:TIM}

Assuming the transverse field Hamiltonian as $H_D$, the total Hamiltonian of QA forms the transverse Ising models (TIMs). We briefly review properties of some representative TIMs in this subsection \cite{kibble_topology_1976,zurek_cosmological_1985,zurek_dynamics_2005,polkovnikov_universal_2005}.

\subsubsection{Pure and disordered transverse Ising chain}\label{sec:1dTIM}

As the simplest case we first consider the pure ferromagnetic  one-dimensional TIM (1dTIM) given by
Eq.~(\ref{eq:pure1dTIM}).
There are two phases of the ground state in this model separated by a quantum phase transition at
$\Gamma/J = 1$. The ground state is disordered for $\Gamma/J > 1$, while it is ferromagnetically ordered for $\Gamma/J < 1$. At the critical point $\Gamma/J=1$, the energy gap above the ground level vanishes as the system size $L\to\infty$. The scaling of the energy gap with the linear size $L$ at the critical point is given by $\Delta\sim L^{-1}$. Critical exponents of the correlation length and the dynamical exponent are $\nu = 1$ and $z = 1$, respectively.

The scaling of the defect density following a  QA of the pure 1dTIM was solved exactly by Dziarmaga \cite{dziarmaga_dynamics_2005}. Let us assume $\Gamma = -tJ/\tau$ with the parameter $\tau$ which denotes the inverse of the annealing speed. Using the periodic boundary condition in Eq.~(\ref{eq:pure1dTIM}), and applying the Jordan-Wigner transformation followed by  the Fourier transformation,  the quantum time evolution of the spin state is reduced to decoupled  Landau-Zener models of two-level systems for each momentum mode \cite{suzuki_quantum_2012,dutta_quantum_2015}. When the time $t$ is varied from $-\infty$ to $0$, with the initial state chosen to be the ground state of the initial Hamiltonian, the residual energy per spin at the final time $t=0$
is found to to be of the form 
\begin{equation}
 \varepsilon_{\rm res} = \frac{1}{\pi}\frac{1}{\sqrt{2J\tau/\hbar}}
\label{eq:Dziarmaga_1dTIM}
\end{equation}
in the thermodynamic limit $L\to\infty$. We recall that the residual energy is defined as the excess energy that is the difference of 
the energy expectation value of $H(t=0)$ with respect to the evolved state at $t=0$ from the
ground energy of $H(t=0)$. According to the Kibble-Zurek scaling with $\nu=z=1$, one has
$\varepsilon_{\rm res}\sim \tau^{-1/2}$ consistent with Eq.~(\ref{eq:Dziarmaga_1dTIM}).

The disordered version of
 1dTIM is given by
\begin{equation}
 H = - \sum_{j}J_j \sigma_j^z \sigma_{j+1}^z - \Gamma\sum_jh_j\sigma_j^x ,
\label{eq:disorder1dTIM}
\end{equation}
where $J_j$ is a random ferromagnetic coupling and $h_i$ is a random transverse field obeying distributions $\pi_J(J)$ and $\pi_h(h)$, respectively. The phase transition of this model happens when $[\log J]_{\rm av} = \log\Gamma + [\log h]_{\rm av}$, where $[\cdots]_{\rm av}$ denotes the random average. 
The ground state is ferromagnetically ordered for $\Gamma < \exp([\log J]_{\rm av} - [\log h]_{\rm av})$, and disordered for $\Gamma > \exp([\log J]_{\rm av} - [\log h]_{\rm av})$ \cite{fisher_random_1992,fisher_critical_1995,young_numerical_1996,fisher_distributions_1998}. 
The phase transition is characterized by the infinite randomness fixed point, where the dimensionless parameter appearing in the distribution of the energy gap $\Delta$  is 
$(-\log\Delta)/\sqrt{L}$, implying that the energy gap  scales as $e^{-C\sqrt{L}}$, where $C$ is a positive constant. Therefore, even though the ground state of this model with $\Gamma=0$ is trivial, the dynamics of QA to this state across the quantum phase transition is highly nontrivial. The size scaling of the typical gap suggests that the time to arrive at the target state by QA scales as subexponential with $L$. In connection to the Kibble-Zurek scaling, Dziarmaga and Caneva et al., reported that the density of kinks produced after QA scales as 
\begin{equation}
 [\rho_{\rm kink}]_{\rm av} \sim 1/\log^2 \alpha\tau
\label{eq:disorder1dTIM_kink}
\end{equation}
with an $\mathcal{O}(1)$ constant $\alpha$ \cite{dziarmaga_dynamics_2006,caneva_adiabatic_2007}. Thus, the defect density decays in a logarithmically slow fashion with the annealing time $\tau$ which we reiterate makes QA difficult. {However, it has been reported that 
SA for the one-dimensional disordered Ising model (i.e., Eq.~(\ref{eq:disorder1dTIM}) with $\Gamma=0$) yields $[\rho_{\rm kink}]_{\rm av} \sim 1/\log \alpha'\tau$, {where $\alpha'$ is a constant}, which decays slower than Eq.~(\ref{eq:disorder1dTIM_kink}) \cite{suzuki_cooling_2009}. Therefore, this model reveals 
an evident advantage of QA over SA.}

\subsubsection{Pure transverse Ising model in higher dimensions}
The two-dimensional TIM (2dTIM) may be the simplest model next to 1dTIM, though unlike the 1d case, the 2d model is not integrable. The equilibrium properties of the 2dTIM has been studied numerically and some of thermodynamic properties including the character of quantum and thermal phase transitions are available. Recently Schmitt et al., carried out numerical study of QA in 2dTIM in the context of the Kibble-Zurek scaling using state-of-the-art numerical methods \cite{schmitt_2021}. Their results are consistent with the Kibble-Zurek prediction. Study 
of {out-of-equilibrium} dynamics of a two-dimensional quantum system will be a direction of study in the near future.

The situation becomes simpler in the infinite dimension. The pure TIM in the infinite dimension is
written as
\begin{equation}
 H = - \frac{J}{N}\sum_{j<k}\sigma_j^z\sigma_{k}^z - \Gamma\sum_j\sigma_j^x ,
\label{eq:pure_infinite-dTIM}
\end{equation}
where $N$ denotes the number of spins. Note that each spin interacts with all the other spins with 
an equal strength. Defining the total spin operator as $\vec{S} = (1/2)\sum_{j=1}^N \vec{\sigma}_j$, this Hamiltonian can be arranged into
\begin{equation}
 H = - 2\frac{J}{N} S_z^2 - 2\Gamma S_x + \frac{J}{2} .
\end{equation}
In the thermodynamic limit, this model undergoes a continuous QPT at $\Gamma = J$. The energy gap above the ground state behaves as $\Delta ~ [(\Gamma - J)\Gamma]^{1/2}$ for $\Gamma \geq J$
and the size scaling of the energy gap at $\Gamma = 1$ is $\Delta_c \sim N^{-1/3}$ \cite{botet_1983}.
Introducing the effective dimension $d_{\rm eff}$ so that the system size $N$ is tied to the linear size $L$ by $L^{d_{\rm eff}}=N$, one has relations between critical exponents as $z\nu = 1/2$ and $z/d_{\rm eff} = 1/3$. Then, assuming $z=1$ as in pure TIMs with finite dimension, one obtains $\nu = 1/2$ and $d_{\rm eff} = 3$. Caneva et al., studied QA of the present model and obtained $\varepsilon_{\rm res}\sim \tau^{-1/3}$. This scaling is inconsistent with the Kibble-Zurek scaling, since the latter predicts $\tau^{-1}$. Acevedo et al., revealed that there is an anomaly in the transition amplitude between the ground and excited states in the present model \cite{acevedo_2014}. Therefore the naive phenomenological argument to derive the Kibble-Zurek scaling does not apply to the system in infinite dimension. We shall also discuss an extension  of the Hamiltonian (\ref{eq:pure_infinite-dTIM}) to the $p$-body interacting model in Sec. \ref{sp_qa} and  argue that the QA does not work in this model with odd $p$.

%\noindent
%\tr{
%---This paragraph will be removed, since $p$-spin model is discussed in Atanu's part.\\
%The Hamiltonian (\ref{eq:pure_infinite-dTIM}) can be extended to the $p$-body interacting model written as
%\begin{equation}
% H = - \frac{J}{N^{p-1}}\sum_{j_1,\cdots,j_p}\sigma_{j_1}^z\cdots\sigma_{j_p}^z - \Gamma\sum_j\sigma_j^x .
%\label{eq:p-TIM}
%\end{equation}
%This model with $\Gamma = 0$ has the doubly degenerated ferromagnetic ground state for even $p$, while for odd $p$ the spin-up state is the unique ground state at $\Gamma = 0$. J\"org et al. showed that this model with odd $p$ involves a first-order QPT and the energy gap at a QPT closes exponentially with the number of spins $N$ \cite{Jorg_2010}. Therefore QA does not work in this model with odd $p$.}

\subsubsection{Transverse Ising spin glass}

The Hamiltonian of the Edwards-Anderson \cite{edwards_theory_1975} version of the transverse Ising spin glass, introduced by Chakrabarti in 1981 \cite{chakrabarti_critical_1981}, is written as
\begin{equation}
 H = -\sum_{\langle j k\rangle} J_{jk} \sigma_j^z\sigma_k^z - \Gamma\sum_j\sigma_j^x ,
\label{eq:EA_model}
\end{equation}
where $\langle j k\rangle$ stands for nearest neighbour pairs and 
$J_{jk}$ are independent random variables. 
The order parameter of a spin glass is defined in terms of the spin overlap between different replicas. Supposing that $\sigma_j^{\alpha,a}$ denotes the spin operator for a replicated system labeled by $a$, the overlap operator between replicas $a=1$ and $a=2$ is defined by
$R_{1,2} = (1/N)\sum_{i=1}^N\sigma_j^{z,1}\sigma_j^{z,2}$. The order parameter is then given by $q = [\langle R_{1,2}\rangle]_{\rm av}$. The spin glass order is characterized by $q > 0$ with zero magnetization $m = 0$, where the magnetization is defined by $m = [\langle (1/N)\sum_{i=1}^N\sigma_j^z]_{\rm av}$. This means that the spin configuration is spatially random but frozen. Rieger et al., and Guo et al., investigated the character of QPTs of this model with the Gaussian distribution of $J_{jk}$ with zero mean and unit variance in square and cubic lattices, respectively, by means of the quantum Monte Carlo simulation \cite{Rieger_1994,Guo_1994}. 
Singh and Young studied $\pm J$ model where $J_{kj}$ takes $+1$ or $-1$ with equal probability for dimensions up to $d = 8$ using the linked cluster expansion to determine the location of the QCP \cite{Singh_2017}.
%Reported values of critical exponents for the Gaussian model were $z \approx 1.5$ and $\nu \approx 1.0 $ for the square lattice and $z \approx 1.3$ and $\nu \approx 0.77$ for the cubic lattice \cite{Rieger_1994,Guo_1994}. 
Subsequently, QPTs of these models were reconsidered by Miyazaki and Nishimori \cite{miyazaki_real-space_2013} and by Matoz-Fernandez and Rom\'a \cite{matoz-fernandez_unconventional_2016} using the real-space renormalization group and 
the quantum Monte-Carlo with parallel-tempering, respectively. They concluded that the QPTs in transverse Ising spin glasses in two and three dimensions were compatible with the infinite randomness fixed point with the critical exponents $\nu$ and $\psi$, where $\psi$ specifies the activation type of size scaling of the energy gap as $[\log\Delta]_{\rm av}\sim N^{\psi/d}$ \cite{motrunich_infinite-randomness_2000,karevski_random_2001,lin_entanglement_2007,igloi_strong_2005,igloi_strong_2018}.
The estimated exponents for the Gaussian model were $\nu \approx 1.2$  and $\psi \approx 0.44$ in two dimension \cite{miyazaki_real-space_2013,matoz-fernandez_unconventional_2016} and $\nu \approx 0.94$ in three dimension \cite{miyazaki_real-space_2013}.

%Reported values of critical exponents were $z \approx 1.5$ and $\nu \approx 1.0 $ for the square lattice and $z \approx 1.3$ and $\nu \approx 0.77$ for the cubic lattice. Singh and Young studied $\pm J$ model where $J_{kj}$ takes $+1$ or $-1$ with equal probability for dimensions up to $d = 8$ using the linked cluster expansion to determine the location of the QCP \cite{Singh_2017}.

The Hamiltonian of the transverse Ising spin glass in infinite dimension, i.e.,  the quantum Sherrington-Kirkpatrick (SK) model,  is written as \cite{Ray_1989}
\begin{equation}
 H = -\frac{1}{\sqrt{N}}\sum_{1\leq j<k\leq N}J_{jk}\sigma_j^z\sigma_k^z
  - \Gamma\sum_{j=1}^N\sigma_j^x ,
\end{equation}
The classical SK model in the absence of the transverse field unveiled the existence of so-called replica symmetry breaking (RSB) in the spin glass phase \cite{parisi1980order,binder1986spin}, where the overlap $R_{1,2}$ has a dispersed continuous distribution in the thermodynamic limit. Ray et al., conjectured on the basis of the quantum Monte-Carlo simulation the collapse of a continuous distribution for the classical SK model into a delta function in the presence of any amount of the transverse field \cite{Ray_1989},
which paved the way for using quantum tunneling in finding the global minimum 
or ground state of SK spin glass model. In the classical model, due to random interactions between spins 
at different lattice sites, such systems have many local minima 
in free energy which are separated by large energy barriers of 
order $O(N)$, where $N$ is the system size~\cite{binder1986spin}. 
This induces non-ergodicity in the system and eventually breaks the 
replica symmetry of the system. As a result, finding the ground state 
or global minimum of such systems is a very hard problem; for 
SK spin glass model, it turns out to be NP (non-deterministic polynomial-time) hard. The system 
indeed gets trapped into one of the local minima inside the 
spin glass phase, due to the highly rugged nature of free-energy landscape. 
This leads to a broad order parameter distribution 
in the spin glass phase\cite{parisi1980order}. 
In addition to a peak value of the order parameter distribution, 
it is extended up to the zero value of the order parameter even 
in the thermodynamic limit.

It seems that the scenario may  change drastically, when a transverse 
field is applied on the SK spin glass \cite{Ray_1989}. The presence of quantum 
fluctuations induces ergodicity in the system, since quantum 
tunneling becomes  possible between the local minima separated by 
tall and narrow free-energy barriers. This indicates the 
restoration of replica symmetry breaking for quantum SK spin 
glass model. As a result, the order parameter distribution 
should be sharply peaked at a point for quantum SK model in 
the thermodynamic limit. This ergodic behavior of quantum SK 
model is responsible for advantage in quantum annealing 
in comparison to simulated annealing.

This conjecture was  criticized by Young \cite{Young_2017} by solving numerically the effective one-dimensional model to which the quantum SK model can be mapped in the $N \to \infty$ limit; this work predicted that the replica symmetric solution is unstable down to zero temperature. On the contrary,  Mukherjee et al., \cite{mukherjee2018possible} explored the behavior of the order parameter distribution of the quantum {SK} model in the spin glass phase using Monte Carlo technique for the effective Suzuki-Trotter Hamiltonian at finite temperatures  
 (see  Eq. \eqref{eq_Trotter} discussed later) and the exact diagonalization method at zero temperature.  It has been found that there exists a low temperature regime 
in the spin glass phase, where the order parameter distribution becomes 
peaked around its most probable value in thermodynamic limit, 
thus suggesting the ergodic behavior. On the other hand, 
the order parameter distribution remains Parisi type in high 
temperature regime, which indicates the non-ergodic behavior 
of the system in this part of the spin glass phase. These 
two regions of the spin glass phase are separated by a boundary, 
connecting the zero temperature-zero transverse field point 
and the quantum-classical crossover point on the phase boundary~\cite{mukherjee2015classical,mukherjee2018possible}. In addition, quantum annealing has also been investigated  for quantum SK model
using Suzuki-Trotter Hamiltonian dynamics in both the ergodic 
and {non-ergodic} regimes. The average annealing time {was estimated},
when both the temperature and the transverse-field {were} annealed 
down to some fixed low values, starting from the paramagnetic 
phase. It {was} found that the average annealing time is 
independent of the system size, when the annealing is performed 
through the ergodic (quantum fluctuation dominated) region, 
whereas it grows strongly with the system size, when the 
annealing is carried out through the non-ergodic (classical 
fluctuation dominated) region. This suggests that the quantum 
annealing has potential to detect whether a phase is ergodic or non-ergodic. Also, the average annealing time to approach a same 
ground state is small for annealing through ergodic regime 
compared to {that through the non-ergodic regime}. The QA for SK spin glass is also 
studied by tuning both transverse and longitudinal fields, and 
it has been shown that this protocol exhibits some effectiveness 
compared to the QA by varying the transverse field only~\cite{rajak2014quantum}.

Recently, Leschke et al., proved rigorously nonzero variance of the overlap in the thermodynamic limit of the quantum SK model at sufficiently low temperature with small but finite transverse field \cite{Leschke_2021}. Their study reveals a dispersed distribution of the overlap, therefore the existence of RSB. However the controversy about the continuous distribution of the overlap in the quantum SK model is still an open problem \cite{schindler2022variational}.

%The quantum SK model has been extended to the $p$-body transverse Ising spin glass,
%\begin{equation}
% H = -\sqrt{\frac{p!}{2N^{p-1}}}\sum_{1\leq j_1 <\cdots < j_p\leq N}
%  J_{j_1 j_2\cdots j_p}\sigma_{j_1}^z\sigma_{j_1}^z\cdots \sigma_{j_p}^z
%  - \Gamma\sum_{j=1}^N\sigma_j^z ,
%\end{equation}
%where $J_{j_1 j_2 \cdots j_p}$ are independent random variables. 

\subsection{Transverse Ising model for satisfiability}

The satisfiability problem is known as one of the basic combinatorial optimization
problem in computer science. Given the number of bits and constraints among bits, 
the problem is to determine whether the bit configuration satisfying all the 
constraints exists or not. In the case of 3-satisfiability problems, or 3-SATs in short,
each constraint involves three bits drawn randomly.
Using the spin language, a constraint for three spins
$S_i$, $S_j$ and $S_k$ taking the values $\pm 1$
can be represented as $(S_i + S_j + S_k - 1)^2$, for instance.
This vanishes and is called satisfied when two of the three spins are $+1$ and
the other spin is $-1$, otherwise it gives a nonzero and positive value.
The Exact Cover, a variant of the 3-SAT, consists of $M$ such 
constraints for $N$ spins, and thus the corresponding quantum model that needs to be
annealed is given by
\begin{equation}
 H = \sum_{m=1}^M (\sigma_{i_m}^z + \sigma_{j_m}^z + \sigma_{k_m}^z - 1)^2
  - \Gamma\sum_{j=1}^N \sigma_j^x .
\label{eq:ExactCover}
\end{equation}
%\tr{Does the first term also contain $N$?}
If one arrives at the exact ground state for $\Gamma = 0$, by annealing the field $\Gamma$, then one can solve the Exact Cover.
However, the Exact Cover is an NP-complete problem which no known algorithm
can solve in a time polynomial in $N$. Young et al., studied Eq.~(\ref{eq:ExactCover})
by means of the quantum Monte-Carlo method \cite{Young_2010}.
They found that some instances of the model show a discontinuous first-order QPT with an 
exponentially small energy gap and the fraction of such instances grows toward unity with
increasing $N$.
J\"org et at. also reported occurrence of a first-order QPT in the random 3-XORSAT
problem, which is another variant of the 3-SAT \cite{jorg_first-order_2010}.

\section{Implementation of quantum annealing}\label{sec:implementation}

Implementing QA is a challenging task, since one needs to evolve a many-spin state
under a quantum many-body Hamiltonian. In this section, we review results
from numerical simulation using real-time dynamics as well as
Monte-Carlo dynamics, and from quantum simulations using hardwares.

\subsection{Results from numerical simulation for real time dynamics}

The real-time evolution of a quantum system governed by the Schr\"odinger equation 
can be computed in general by solving a linear differential equation. 
However, since the number of unknown functions of time
increases as $2^N$ with the number of spins, the system size acceptable to a
conventional computer is limited to $N\sim O(10)$. 
Kadowaki and Nishimori reported in the seminal paper that 
QA yields better solutions than the classical simulated annealing
on the basis of their simulation for 8 spin systems of a frustrated model and the SK model
\cite{kadowaki_quantum_1998}. Farhi et al.,  reported the numerical result
for the Exact Cover problem with the number of spins up to 20 \cite{farhi_quantum_2001}.
Exact Cover is one of the NP-complete problems, which no known classical algorithm
can solve in a time polynomial in the number of spins. The numerical result
suggested a quadratic scaling of the runtime in QA with respect to the number of spins.
However, this scaling should turn into an exponential one for larger size as
shown by the quantum Monte-Carlo study for the energy gap \cite{Young_2010}.

Restricted to systems in one dimension, there are efficient methods of
numerical simulation for real-time evolution. In case of 1dTIMs without 
the longitudinal field, irrespective of disorder, the Schr\"odinger
equation of the spin state reduces to the Bogoliubov-de Gennes equation
of $2N$ unknown functions of time through the Jordan-Wigner's fermionization and
the Bogoliubov transformation \cite{barouch_statistical_1970}.
Generic 1dTIMs with longitudinal fields cannot be mapped to free fermion
models. However, time evolution of generic 1dTIMs 
can be simulated using 
the time-dependent density matrix renormalization group (tDMRG) 
proposed by White and feiguin \cite{white_real-time_2004}
or the time evolving block decimation (TEBD) by Vidal \cite{vidal_efficient_2004}.
In addition, the infinite system of 1dTIMs with no disorder can be simulated
using an infinite method of TEBD (iTEBD) \cite{orus_infinite_2008}.
These methods serve the study of QA in 1dTIM with
a uniform or disordered longitudinal field \cite{pollmann_dynamics_2010,king_coherent_2022}.

\subsection{Results from numerical simulation for Monte-Carlo dynamics}

\begin{figure}[t]
\begin{center}
 \includegraphics[width=10cm, bb = 0 0 720 354]{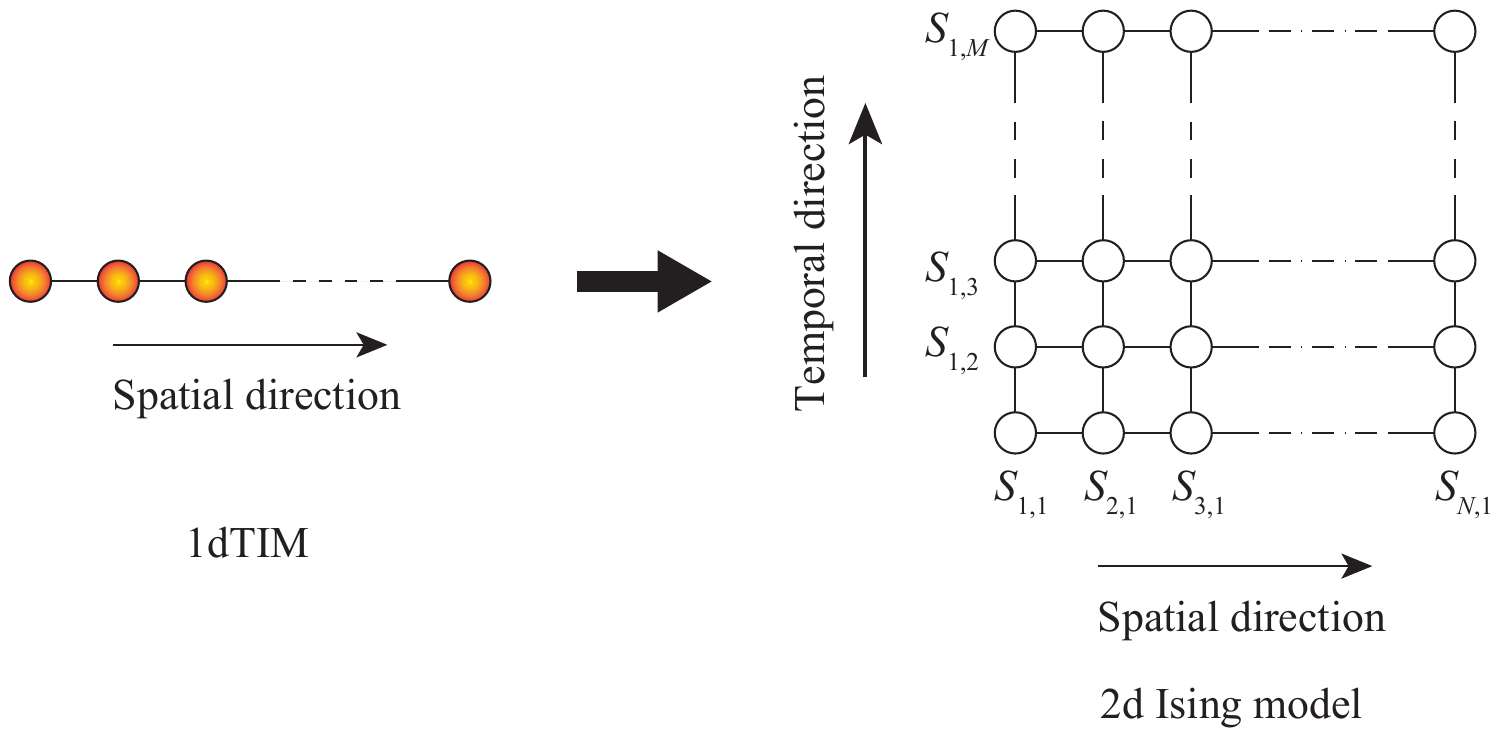}
\end{center} 
\caption{Schematic picture of the Suzuki-Trotter mapping. A 
1dTIM is mapped to a two-dimensional classical Ising model on the square lattice. The
additional dimension corresponds to the time. $S_{j,m}$ denotes an Ising-spin
variable at spatial site $j$ and temporal site $m$.} 
\label{fig:Suzuki_Trotter}
\end{figure}
A $d$-dimensional TIM in finite temperature (denoted by $\beta^{-1}$) 
with spin-spin coupling $J_{jk}$ and transverse field $\Gamma$ can
be mapped to a $(d+1)$-dimensional classical Ising model by the Suzuki-Trotter mapping
\cite{Trotter_product_1959,suzuki_quantum_1986}.
The resulting Hamiltonian $H_{\rm eff}$ is given by
\begin{equation}\label{eq_Trotter}
 H_{\rm eff} = -\sum_{m=1}^M\sum_{j<k}\frac{J_{jk}}{M}S_{j,m}S_{k,m}
  - { \frac{M}{2\beta}\log\coth\frac{\beta\Gamma}{M}}\sum_{j=1}^N\sum_{m=1}^M
  S_{j,m}S_{j,m+1} ,
\end{equation}
where $\beta$ is the inverse temperature, $M$ is the Trotter number,
$S_{j,m}$ denotes the spin variable with the spatial site $j$ and
temporal site $m$ taking values $\pm 1$, and we defined the sign of $J_{jk}$ according to 
Eq.~(\ref{eq:EA_model}). In Figure \ref{fig:Suzuki_Trotter}, we schematically illustrate the mapping
of 1dTIM into a (1+1)-dimensional classical Ising model.
One can simulate in principle any TIM in and
out of equilibrium using this effective Hamiltonian and the Monte-Carlo
method. This method is called the quantum Monte-Carlo method (QMC).
Although the number $\beta/M$ controls the accuracy of QMC, the cluster-update
method invented by Swendsen and Wang along the temporal direction
enables to have $\beta/M\to 0$ \cite{swendsen_nonuniversal_1987,nakamura_quantum_2003}.
QMC is known to give rise to the sign problem and fail when the model involves
the frustration. However, QMC for TIM is free from the sign problem. Therefore
QMC is a powerful method of classical computation in simulating TIM.

QA can be implemented
in QMC by regarding the Monte-Carlo step as time. The dynamics
realized by QMC is not the quantum dynamics governed by the Schr\"{o}dinger
equation but the stochastic one. However, QA with QMC serves the purpose of
solving an optimization problem using a classical computer \cite{kadowaki_arxiv}.
Several works have shown so far that QA with QMC works in variety of optimization problems, such as
two-dimensional Ising spin glass \cite{santoro_theory_2002,martonak_quantum_2002}, travelling salesman problem \cite{martonak_quantum_2004}, and 3-SATs \cite{battaglia_optimization_2005}.
\begin{figure}[t]
 \begin{center}
  \includegraphics[width=8cm, bb = 0 0 232 178]{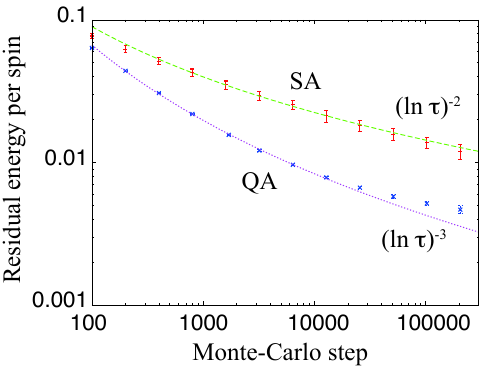}
 \end{center}
\caption{Comparison of the residual energy between QA by QMC and SA for the two-dimensional spin glass model with 99$\times$99 spins with random coupling $J_{jk}$ from the uniform distribution between -2 and 2. The cluster-flip algorithm in the imaginary-time direction was used in QMC. SA was started from the initial temperature $T=5.0$, while QA was from $\Gamma = 5.0$ with $T=0.01$. The average was taken over 100 runs for a single instance in SA, and for 16 instances in QA. The decay of the residual energy in SA is well fitted by $(\log \tau)^{-2}$. As for QA, it is approximated by $(\log\tau)^{-3}$, except for long annealing time where the decay rate is smaller. (Taken from ref.~\cite{suzuki_kibble-zurek_2011}.)}
\label{fig:QMC_spinglass2D}
\end{figure}
Figure \ref{fig:QMC_spinglass2D} shows comparison between QA by QMC and SA in the two-dimensional
spin glass model \cite{suzuki_kibble-zurek_2011}. This result implies outperformance of QA over SA.
Although an opposite result has been reported for harder 3-SAT problems \cite{battaglia_optimization_2005}, numerical studies using QMC suggest that there are problems for which QA is potentially advantageous over SA due to the restoration of ergodicity by quantum fluctuation \cite{Ray_1989}.

\subsection{Results from quantum simulation using hardware}

The most efficient way to perform QA is use a quantum magnetic material which
realizes a TIM with a temporally controllable transverse field.
LiHo$_x$Y$_{1-x}$F$_4$ is a material that models a disordered Ising model
and it also realizes a disordered TIM by application of the magnetic field
perpendicular to the easy axis of magnetization \cite{wu_classical_1991}. 
Brooke et al.,  investigated two protocols, QA and thermal annealing (TA), 
using this material with $x=0.44$.
In QA, the transverse field $\Gamma$ was strengthened at high temperature, the system
was cooled, and then $\Gamma$ is weakened to $\Gamma_f$ at low temperature.
On the other hand, in TA, the temperature was lowered with keeping $\Gamma = 0$, and
then $\Gamma$ was raised to $\Gamma_f$. In both protocols, the initial and final sets
of $\Gamma$ and temperature were the same and the durations were the same as well. 
The ground state at $\Gamma_f$ as the target
state is a glassy ferromagnetic state. 
Brooke et al., reported that the state after QA much is closer
to an equilibrium state than the state after TA \cite{brooke_quantum_1999}. 
This result implies that QA brings the state to the target faster than TA.
Quite recently S\"aubert et al., studied QA and TA of the same material 
and detailed the dynamical behavior of the energy landscape during QA.
They showed that the transverse field applied in QA induced random longitudinal
fields, implying that the energy landscape of the problem Hamiltonian $H_P$ evolved as QA proceeded \cite{saubert_microscopics_2021}. This evolving landscape may
be an issue of future work related to QA.

Progress in Rydberg atom experiments enables to use Rydberg atoms as
a quantum simulator. Keesling et al., performed a quantum simulation of a sort of QA
using an array of 51 Rydberg aroms. In this simulation, the system is described
by the many-body Hamiltonian $H=(\Omega/2)\sum_i(|g_i\rangle\langle r_i| + h.c.)
- \Delta\sum_i n_i + \sum_{j<k}V_{jk}n_j n_k$,
where $|g_j\rangle$ and $|r_j\rangle$ denote the ground and the excited Rydberg states, respectively, of atoms, $n_j= |r_j\rangle\langle r_j|$, and $V_{jk}$ is the van der Waars interaction with the strength which decays as $1/|j-k|^6$.  The model exhibits
a QPT, achieved by tuning the parameter $\Delta$, belonging to the same universality class as that of 1dTIM . Keesling et al.,  observed
that the Kibble-Zurek scaling for the kink density arising due to the sweeping 
of the parameter $\Delta$, turns out to be the very same as that in 1dTIM \cite{keesling_quantum_2019}.

In order to apply QA as a computation to an optimization problem in practice, 
spin-spin interactions and longitudinal fields in addition to the transverse field
need to be locally controllable. A Canadian venture company, D-Wave Systems, 
has developed a quantum annealing machine named as a quantum annealer, which
consists of programmable coupled superconducting flux qubits and 
performs QA to various Ising models \cite{johnson_quantum_2011}. 
The number of qubits in the latest machine is beyond five thousands.
This is 100 times larger than the number of qubits in the current gate-based
quantum computer. 
Denchev et al., benchmarked D-Wave 2X using 100 instances of the weak-strong cluster model
with up to 945 spins \cite{denchev_what_2016}. Qubits in D-Wave 2X form the so-called chimera graph with unit cells consisting of 8 qubits. In the weak-strong cluster model, there are all-all ferromagnetic couplings inside the cell, and half of the spins in a cell ferromagnatically couple with those in neighboring cells. In addition, weak longitudinal fields are applied to spins in randomly chosen cell, while strong fields anti-parallel to the weak ones are applied to spins in the other cells.
\begin{figure}[t]
 \begin{center}
  \includegraphics[width=8cm, bb=0 0 242 189]{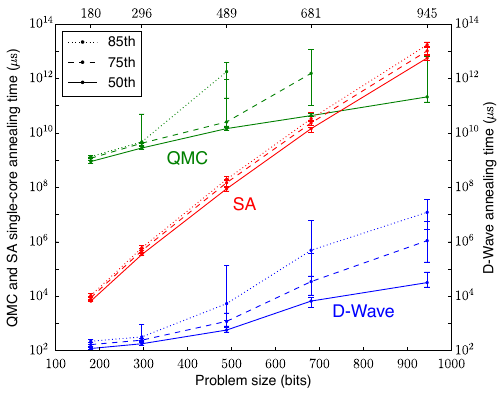}
 \end{center}
 \caption{Comparison of the time to reach the ground state with 99\% success probability as a function of the problem size in D-Wave 2X, simulated annealing (SA), and quantum Monte-Calro (QMC). The runtime in SA and QMC is defined by $n_{\rm sweeps}NT_{\rm update}$, where $n_{\rm sweeps}$ is the number of sweeps (one sweep attempts to update all spins). $T_{\rm update}$ is the single-spin update time for SA and the update time of a spin-cluster along the temporal direction. It is set as $T_{\rm update}=1/5$ ns for SA and $10\times 870$ ns for QMC. Data for 50th, 75th, and 85th percentile taken from a set of 100 instances are shown. The error bars represent 95\% confidence interval from bootstrapping. Taken from ref.~\cite{denchev_what_2016}.}
 \label{fig:Denchev2016}
\end{figure}
Figure \ref{fig:Denchev2016} shows the time to reach the ground state with 99\% success probability. 
For D-Wave, this time is defined by 20 $\mu$s $[\log (1 - 0.99)/\log (1 - p)]$
for an instance, where
the annealing time is fixed at 20 $\mu$s and $p$ denotes
the success probability to obtain the ground state estimated from many runs. 
As for SA and QMC, it is the runtime
on a single processor. Regarding the median from 100 instances, D-Wave 2X is 10$^8$ and 10$^7$ times faster than SA and QMC, respectively. 
%We remark that the weak-strong cluster model is a hard problem to be solved by SA, since spins form clusters in this model whereas each spin flips one by one in SA. Therefore this result does not necessarily imply that D-Wave' quantum annealer is much faster than SA for any other optimization problems.

\begin{figure}
\begin{center}
 \includegraphics[width = 12cm, bb = 0 0 325 128]{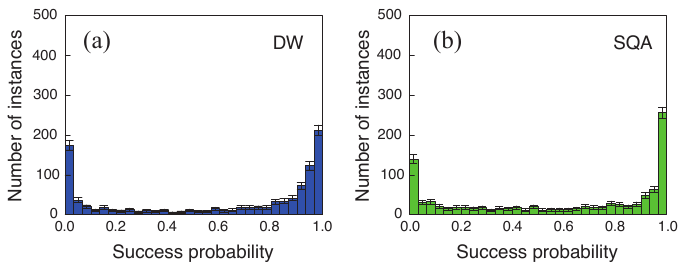} 
\end{center}
 \caption{Comparison of the histogram of the probability of finding the ground state between D-Wave One and QA with QMC (simulated QA) over 1000 instances of the spin glass model with $N=108$ spins. Taken from ref.~\cite{boixo_evidence_2014}.}
 \label{fig:Boixo2014}
\end{figure}

Boixo et al., tested D-Wave's quantum annealer to a spin glass
model $H_P = - \sum_{\langle j k\rangle}J_{jk}\sigma_j^z\sigma_k^z$, where
$J_{j k}$ is chosen randomly from $J=\pm 1$, with $N=108$ spins and reported
that the results of quantum annealer correlated well with those obtained by
QA with QMC \cite{boixo_evidence_2014}. Figure \ref{fig:Boixo2014} shows the comparison
of the histogram of the success probability between D-Wave's quantum annealer (DW) and QA with QMC (named as Simulated QA). The bimodal distribution which is common in D-Wave and QMC could be an evidence that the system embedded in D-Wave's quantum annealer was a quantum system. 
However, Shin et al., reported that the classical spin vector model 
 along with
the Monte-Carlo dynamics, named as the spin vector Monte-Carlo (SVMC) model, 
provided as strong correlation with D-Wave's data as QMC
\cite{shin_how_2014}. The classical spin vector model is represented by the Hamiltonian
\begin{equation}
 H(t) = -A(t)\sum_{j=1}^N \cos\theta_j - B(t)\sum_{\langle j k\rangle}J_{jk}
  \sin\theta_j\sin\theta_k ,
\end{equation}
where $\theta_j$ denotes the angle of the unit vector at site $j$ in the $xz$ plane.
These works have raised a problem to identify the model of D-Wave's quantum annealer 
\cite{suzuki_quantum_2015}.
Bando et al.,  studied the Kibble-Zurek scaling in 1dTIM using D-Wave 2000Q and
found that the kink density defined by 
$n=(1/2N)\sum_{j=1}^N (1 - \langle\sigma_j^z\sigma_{j+1}^z\rangle) = \varepsilon_{\rm res}/2$ 
scaled with the annealing time $t_a$ as $n\sim t_a^{-\alpha}$ with $\alpha\approx 0.20$
by the device at NASA and $\alpha\approx 0.34$ by the one at D-Wave Systems
\cite{bando_probing_2020}. As mentioned in Sec. \ref{sec:1dTIM}, the scaling of the
kink density is predicted as $n\sim t_a^{-1/2}$ for an isolated system belonging to
one-dimensional Ising universality class.
The authors in ref.~\cite{bando_probing_2020} compared numerical simulations
for 1dTIM with coupling to an environment and for SVMC with the experiment, 
and concluded that the quantum model agreed better with the experiment.
Recently, King et al., studied QA of 1dTIM using D-Wave 2000Q \cite{king_coherent_2022}, 
focusing on shorter annealing times than those in the previous works. 
For short annealing times, the system in the device is less affected by environment as we
shall discuss in the next section.
Comparing analytic and numerical computation for the Schr\"odinger dynamics of the
isolated 1dTIM, the QMC simulation, and the SVMC simulation with the experiment
by D-Wave 2000Q, King et al.,  reported that only the Schr\"odinger dynamics
of the isolated 1dTIM with a small amount of disorder can explain all the
experimental results \cite{king_coherent_2022}.
Also, in ref.~\cite{venturelli2015quantum}, fully connected Sherrington-Kirkpatrick model with random
couplings was programmed using D-Wave ${\rm Two}^{\rm TM}$  annealer, where optimal parameter setting allowed better
performance of the quantum annealer when compared to those obtained using optimized simulated
annealing algorithms.

%\section{Experimental study}

\section{Effects of environment on QA}

\label{Sec:environment}

Although QA is ideally performed in an isolated system, any real system is always coupled
to an  environment and hence susceptible to decoherence. In fact, the system in D-Wave's device is believed to
be affected considerably by an environment when the annealing time is longer than
a few $\mu$s. Therefore it is very important to study an effect of
environment in QA.

%\subsection{Modelling the environment}
There is a variety of models representing an environment. 
Caldeira and Leggett, in their seminal work analyzed the dynamics of flux state
in a SQUID and constructed a simple model of a two-level system coupled to a boson bath, where
bosons are attributed to the electro-magnetic field coming from the fluctuating current
\cite{caldeira_quantum_1983}.
Leggett et al.,  elaborated on the single-spin model coupled a boson bath
\cite{leggett_dynamics_1987}.
Thus, considering QA performed with superconducting flux qubits, the model
which includes the effect of an environment should be an extension of the
Caldeira-Leggett model to many spins.  Hamiltonian is written as
$H(t) = H_S(t) + H_B + H_{\rm int}$, where the system Hamiltonian $H_S(t)$ is given by
Eq.~(\ref{eq:QA_Hamiltonian}). The bath is represented by the collection of harmonic oscillators. Hence the bath Hamiltonian $H_B$ is given, using the boson
operators $b_{j,a}$ and $b_{j,a}^{\dagger}$ for site $j$ and mode $a$, by
\begin{eqnarray}
 H_B = \sum_{j,a}\hbar\omega_a b_{j,a}^{\dagger} b_{j,a} ,
\label{eq:H_B}
\end{eqnarray}
where $\omega_{a}$ is the frequency of the harmonic oscillator of mode $a$.
$H_{\rm int}$ represents the interaction between the system and the bath, and written as
\begin{equation}
 H_{\rm int} = \sum_{j=1}^N \left(\sigma_j^x Q_j^x
  +  \sigma_j^z Q_j^z\right)
 \label{eq:H_int}
\end{equation}
where $Q_j^{\gamma}$ ($\gamma = x$, $z$) is the bath operators given by
$Q_j^{\gamma} = \sum_a\lambda_a^{\gamma}(b_{j,a}^{\dagger} + b_{j,a})$.
The spectral density of the boson bath is assumed as
$J_{\gamma}(\omega) = \sum_a(\lambda_a^{\gamma})^2\delta (\omega - \omega_a)
= \eta_{\gamma} \omega^s e^{-\omega/\omega_c}$, where $\eta_{\gamma}$ denotes the coupling
strength of the system-bath interaction and $\omega_c$ is the energy cutoff
of the bath spectrum. The Ohmic bath refers to $s=1$, while the super-Ohmic and 
sub-Ohmic baths refer to $s > 1$ and $s < 1$, respectively.

%\subsection{Methods}
Let us now move to study the time evolution assuming that  the state of the composite system is
described by the density operator $\rho(t)$ at the instant $t$. The initial state is assumed to be a direct product state of the form 
$\rho(0)=|\psi(0)\rangle\langle\psi(0)|\otimes e^{-H_B/T}/Z_B$, where
$|\psi(0)\rangle$ denotes a state vector of the system, $T$ is the temperature, 
and $Z_B$ is the partition function of the bath. 
Since we are interested in the behavior of the system, we consider
the reduced density operator describing the system, $\rho_S(t) = {\rm Tr}_B \rho(t)$, where
${\rm Tr}_B$ stands for the trace with respect to the bosonic degrees of freedom.

Amin explored the success probability of QA for a range of annealing time $t_a$
obtained by solving numerically the quantum Redfield master equation for
an instance of 16 spins of random Ising model in a random longitudinal field
with nonzero $\eta_z$ and $\eta_x=0$ \cite{amin_searching_2015}. 
The obtained success probability 
is a nonmonotonic function of $t_a$. For short $t_a$, the spin system
is not influenced by the bath and hence the success probability increases
with increasing $t_a$. In a middle range of $t_a$, the thermal environment
disturbs the system's adiabatic evolution more for longer $t_a$, hence leading
to decreasing success probability. For very long $t_a$, finally, the system
evolves keeping the thermal equilibrium with the bath until it is frozen near
the end of QA. The freezing happens because 
$H_S(t_a)$ and $H_{\rm int}$ are commutable when $\eta_x=0$
and hence the relaxation time diverges as $t\to t_a$.
Thus the success probability in this regime goes to the probability of the
ground state at the thermal distribution as $t_a\to\infty$.
%This freezing picture suggests that QA in the thermal environment
%with sufficiently long $t_a$ should produce a thermal distribution of 
%$H_S(t^{\ast})$ with $t^{\ast}$ being the time at which the system is frozen.
%Taking into account $1-t^{\ast}/t_a\ll 1$ for long $t_a$, 
%$H_S(t^{\ast})$ is approximated as $H_S(t^{\ast})\approx B(t^{\ast})H_P$.
%Hence this distribution is approximately the thermal distribution of 
%$H_P$ with the temperature $T/B(t^{\ast})$.

QA of 1dTIM in the presence of an environment has been attracted a lot
of attention in the context of the Kibble-Zurek scaling.
Assuming $Q_j^z=0$, namely, the boson bath coupled to $\sigma^x$,
1dTIM with coupling to the boson bath is mapped to a noninteracting 
fermion model with a fermion-boson coupling through the Jordan-Wigner
transformation. Then the problem is significantly tractable, compared to the
situation with $\eta_z\neq 0$. Patan\'e et al., studied the density of 
excitation following QA using the Keldish technique. Based on the
ansatz that the density of excitation $\mathcal{E}$ is given by 
the sum of the coherent part $\mathcal{E}_{\rm coh}$ 
and the incoherent part $\mathcal{E}_{\rm inc}$ due to the environment,
Patan\'e et al., obtained $\mathcal{E}_{\rm inc}\sim \eta_xT^4\tau$
for the Ohmic bath with temperature $T$ when QA ends near the quantum critical point
\cite{patane_adiabatic_2008,patane_adiabatic_2009}.
The incoherent part increases with $\tau$ in contrast to the coherent part,
hence its scaling is called the anti-Kibble-Zurek scalng.
Nalbach et al., studied the model with the spatially correlated bath
where all spins are coupled to a single bath, i.e.,
$H_{\rm int}= Q^x\sum_j\sigma_j^x$ and $H_B = \sum_a\hbar\omega_z b_a^{\dagger}b_a$
with $Q = \sum_a\lambda_a(b_a^{\dagger} + b_a)$.
In this situation, the correlation length is the largest of the 
Kibble-Zurek length scale $\xi_{\rm KZ}\sim \tau^{1/2}$ and
the thermal length scale $\xi_T\sim T^{-1}$. When $1 \ll \xi_T < \xi_{\rm KZ}$,
the thermal effect comes into play in the density of excitation.
Thus, it is suggested that
$\mathcal{E}\sim \eta T(\tau T^2)\sim \eta_x T^3\tau$
for $\sqrt{\tau}\ll T\ll 1$ in this model.
Nalbach et al., proposed this scaling relation and confirmed using the
dissipative Landau-Zener theory for the two-level system
\cite{nalbach_quantum_2015}. Dutta et al.,  studied 1dTIM in the presence of 
a spatially homogeneous Gaussian white noise on the transverse field, 
instead of considering the coupling to a boson bath.
This stochastic perturbation 
yields an effective dynamics of the noise-averaged density operator
for an open quantum system.
Then the noise-averaged density of excitation has an incoherent part
which scales as
$\mathcal{E}_{\rm inc}\sim \tilde{\eta}_x \tau$, where $\tilde{\eta}_x$ is the strength
of the noise \cite{dutta_anti-kibble-zurek_2016}.
Weinberg et al.,  studied a similar model but with a spatially uncorrelated and temporally 
correlated noise. The numerical result showed the same scaling as ref.~\cite{dutta_anti-kibble-zurek_2016}.
Weinberg et al., also performed quantum simulation for 2dTIM using D-Wave 2000Q and
obtained scaling for the residual energy
$\varepsilon_{\rm res}\sim a\tau^{-\alpha} + b\tau^{\beta}$ with
$\alpha\approx 0.74$ and $\beta \approx 0.456$ \cite{weinberg_scaling_2020}. 
The first term is consistent with
the Kibble-Zurek scaling for 2dTIM with the exponent $d\nu/(z\nu + 1)\approx 0.77$, where
$d =2$, $z=1$ and $\nu \approx 0.63$ \cite{hasenbusch_critical_1999}.
Note that the scaling of the second term corresponding to the incoherent part 
is different from that in 1dTIM, implying that $\mathcal{E}_{\rm inc}\sim \tau$ is specific
to 1dTIM. The anti-Kibble Zurek mechanism in the presence of non-thermal bath
has been also discussed in the framework of the Lindblad formalism
in refs.~\cite{karl_strongly_2017,keck_dissipation_2017,bandyopadhyay_dynamical_2020,rossini_dynamic_2020,puebla_universal_2020}

The Kibble-Zurek and anti-Kibble-Zurek scalings
implies the existence of a global or local minimum of the density of excitation.
Based on the numerical study using the Redfield master equation
in the momentum space
for 1dTIM with the boson bath, Eqs. (\ref{eq:H_B}) and (\ref{eq:H_int}) with $Q_j^z=0$,
Arceci et al., identified the region in the $T-\eta_x$-plane
where there exists the local or global minimum of the density of defects after QA
\cite{arceci_optimal_2018}. Interestingly, a global minimum of the density of defects
appears in the digitized QA, in which the time-evolution operator inducing QA is
split  into slices with a finite time width and further split into
those involving only $H_D$ and those involving only $H_P$ \cite{mbeng_optimal_2019},
suggesting that the decomposition of the time-evolution operator has an influence
as a decoherence on the dynamics of a closed system.

Many studies of 1dTIM with a boson bath focusing on the Kibble-Zurek
physics have assumed $Q_j^z = 0$, namely, coupling between the system operators $\sigma_j^x$ and bosons.
However, in experimental systems such as those made of coupled superconducting flux qubits,
coupling between $\sigma_j^z$ and $Q_J^z$ is rather important \cite{leggett_dynamics_1987}.
Recently, Suzuki et al., developed new 
matrix-product-state-based methods for 1dTIM with
a boson bath with $H_{\rm int}=\sum_j Q_j^z\sigma_j^z$, which enables 
the simulation of finite pure and disordered systems with $O(10^2)$ spins \cite{suzuki_quantum_2019,oshiyama_kibblezurek_2020} or an
infinite translationally invariant system \cite{oshiyama_classical_2022}.
Using the infinite-system method, Oshiyama found modified Kibble-Zurek scaling in 1dTIM
coupled to the bath at zero temperature \cite{oshiyama_kibblezurek_2020}.
Oshiyama et al.,  also studied QA of 1dTIM with the bath at finite temperatures.
In the thermal environment at finite temperature $T$, the infinitely slow QA 
($t_a\to\infty$) can be regarded as the quasistatic and isothermal process, hence
the final energy should be identical to the thermal average of $B(t_a)H_P$ at $T$.
When $t_a$ is sufficiently long but finite, the energy of the system has an excess
from the thermal average. Oshiyama et al.,  found and numerically confirmed that this excess
energy scales with $t_a$ as $t_a^{-1/3}$, for the linear annealing protocol\cite{oshiyama_classical_2022}.

%Rossini and Vicari \cite{rossini_dynamic_2020}

\section{\tb{Avoiding first-order phase transitions: Closed systems}}
\label{sp_qa}
As discussed in Sec.~\ref{sec:Introduction}\ref{sec:QPT_QA}, it has been generally 
observed (with a few exceptions) that the minimum energy 
gap decreases exponentially with the system size for a 
first-order phase transition, whereas it shows a polynomial decrement in 
system size for a second-order phase transition. Therefore, the order of 
phase transitions is an important factor in determining the efficiency 
of the quantum annealing algorithm.

\subsection{Quantum ferromagnetic model}
\label{pspin_model}
We consider here a ferromagnetic $p$-spin model in the transverse field. 
The Hamiltonian for such a system is given by
\begin{equation}
 H = -N\Bigl(\frac{1}{N}\sum _{i=1}^{N} \sigma_{i}^{z}\Bigr) ^{p}-\Gamma\sum_{i=1}^N\sigma_i^x,
 \label{p-spin model}
\end{equation}
where $\sigma_i^z$ and $\sigma_i^x$ are usual Pauli matrices at the lattice site $i$, $\Gamma$ is the 
magnetic field in transverse direction and $N$ is the number of spins and {$p$ is an integer}. These type of models were initially introduced 
in the context of spin glasses. The ground state of the classical model {at zero temperature} with  $\Gamma=0$, corresponds to all 
spins aligned in the same direction. For even $p$, all the spins in up or down states are valid ground 
states, whereas odd $p$ has a unique ground state when all the spins are in up states. Therefore, for 
simplicity, we will concentrate here on the odd $p$ cases. For $p=2$, the Hamiltonian in Eq.~(\ref{p-spin model}) 
reduces to {an infinite-range Ising model which can be mapped to}
 the usual {mean field} Curie-Weiss model exhibiting  continuous phase transitions. On the other hand, for $p>2$, 
both classical and quantum phase transitions of the system are discontinuous.

\begin{figure}[t]
\centering
  \includegraphics[width=6cm, bb = 0 0 360 252]{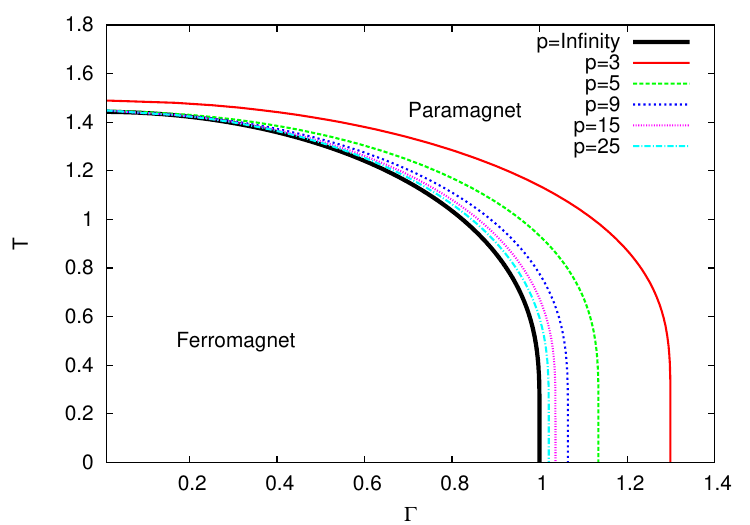}
\caption{Phase diagram of the ferromagnetic $p$-spin model in the $T-\Gamma$ plane for different values of $p$. The ferromagnetic 
and quantum paramagnetic phases are separated by first-order phase 
transitions.
%A first-order transition separates the
%ferromagnetic and quantum paramagnetic phases. 
    Taken from~\cite{jorg2010energy}.}
  \label{pspin_phase_diagram}
\end{figure}

Using Suzuki-Trotter formalism and ``static'' approximation, the phase diagram of the $p$-spin ferromagnetic model 
can be found in $\Gamma-T$ plane for different values of $p$ (see Fig.~\ref{pspin_phase_diagram})~\cite{jorg2010energy}. 
%\tr{In the limit 
%of low-temperature $T$, the first excited state of the system is given by $E=E_{\rm GS}+\Delta Ee^{-\beta\Delta E}$, 
%where $\Delta E=2\sqrt{\Gamma^2 + p^2m ^{2p-2}}$, being the energy gap and $\beta$ is the
%inverse temperature. This indicates that the energy gap between 
%the ground and excited states closes exponentially fast with the system size at the transition} \tr{what is
%$m$? the exponential dependence with the system size is not clearly visible}.
%\tb{Adding the part below by deleting the red part above.}
{In the limit of $p\rightarrow\infty$, using perturbation theory, the minimum energy gap of the system can be calculated as 
$\Delta_{\rm min}=2N2^{-N/2}$~\cite{jorg2010energy}. This indicates that the energy gap between the ground and excited 
states closes exponentially fast with the system size at the transition point. For a general $p$,  an explicit form of 
the energy gap is not available so that one can comment about its scaling with the system size, however, the same 
scaling can be inferred  from numerical calculations.}

\begin{figure}[t]
\includegraphics[width=6cm, bb = 0 0 360 252]{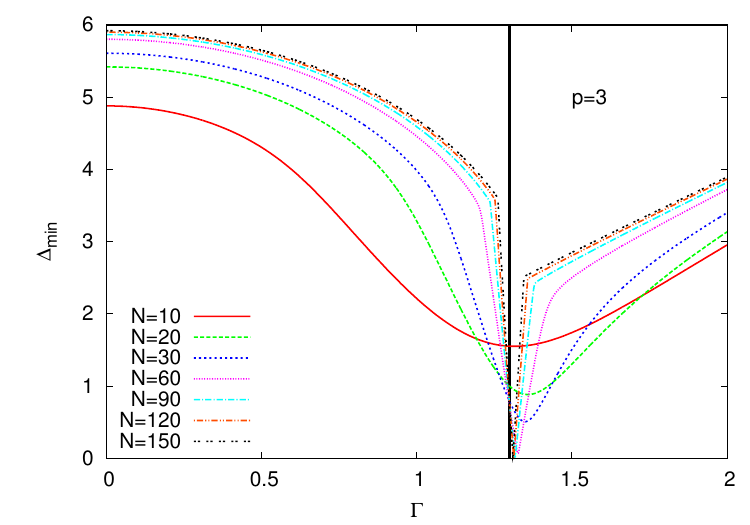}
  \includegraphics[width=6cm, bb = 0 0 360 252]{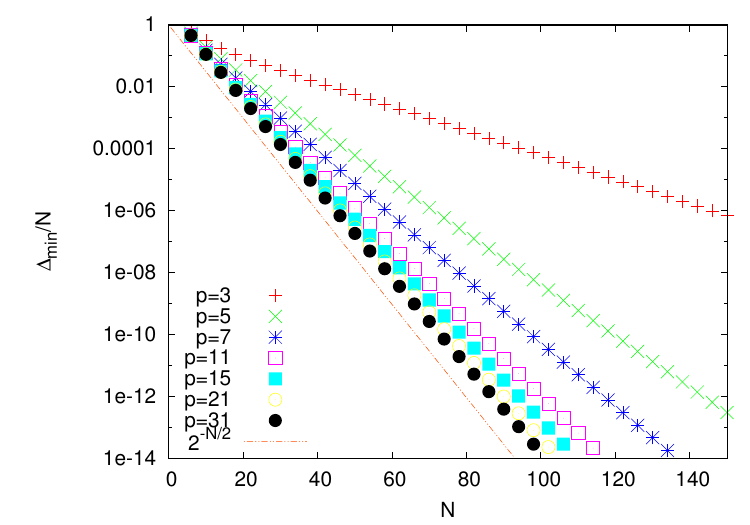}\\
  \caption{Left Panel: Variation of the energy gap as a function 
  of $\Gamma$ for $p=3$ computed using an exact diagonalization method as described in the text. The gap vanishes exponentially 
  fast with the system size $N$ near the critical point $\Gamma_c$ 
  (the black vertical line). It also shrinks near the critical 
  regime as $N$ increases. 
  Right Panel: Minimum energy gap as a function of $N$ for a few 
  values of $p$ on semi-log scale. It shows an exponential fall 
  of the minimum gap with $N$ for all values of $p$ according 
  the scaling relation $\Delta_{min} \propto N 2^{-N\alpha}$.
  Taken from~\cite{jorg2010energy}.
    \label{gaps_scaling}}
\end{figure}

The energy gap of the system can be calculated numerically using two complementary methods as a function of the 
transverse-field $\Gamma$~\cite{jorg2010energy}. Using these numerical methods, we can find the transition point $\Gamma_c$ where the 
energy gap shows a minima that scales with the system size. In the present case, the energy spectrum of the system 
has been studied for $3\le p\le31$. The Hamiltonian in Eq.~(\ref{p-spin model}) is represented by a sparse 
matrix of dimension $2^N$. For such systems, Lanczos method provides nearly exact extreme eigenvalues of the 
Hamiltonian for the system size $N\le21$. From the results of the Lanczos method for $N\le21$, it has been found that 
the transition happens between two states with the maximum possible angular momentum $l=N/2$. The efficiency of 
the numerical simulation can be improved by exploiting the fact that the total angular momentum $L^2$ commutes with the Hamiltonian $H$ 
in Eq.~(\ref{p-spin model}), {(where $L$  is the total angular momentum of $N$ spins)}.  Therefore the transition 
occurs mainly in the subspace of dimension $2l+1=N+1$. In this subspace, 
the Hamiltonian assumes a tri-diagonal form and the resulting tri-diagonal matrix can be diagonalized efficiently 
for a system with size $N\sim100$ in just a few seconds. The energy gap has been shown in the left panel 
of Fig.~\ref{gaps_scaling} as a function of $\Gamma$ for $p=3$ and different $N$. The gap becomes minimum 
at the critical value of $\Gamma$ that agrees with analytically predicted value. One can observe that 
the region where the gap closes gets narrower as the value of $N$ is increased. The minimum energy gap 
$\Delta_{\rm min}$ is further plotted as a function of $N$ for different values of $p$ to find its 
dependence on $N$ (see right panel of Fig.~\ref{gaps_scaling}). It has been found that the minimum energy 
gap decays exponentially as $\Delta_{\rm min}\varpropto N2^{-N\alpha}$ for $p\ge3$. The minimum energy 
gap closes exponentially fast as expected for the first order phase transition. The value of exponent 
$\alpha$ can be computed numerically from the right panel of Fig.~\ref{gaps_scaling}. These exponents 
are also calculated analytically using instantonic approach. A comparison of values of $\alpha$ for 
different values of $p$ are given in Table $1$ of Ref.~[\cite{jorg2010energy}].

Due to an exponential decay of energy gap with the system size, the running time increases exponentially for the 
case of a first-order phase transition, and thus reducing the efficiency of QA process. Therefore, it is an important issue 
to investigate whether one can avoid first-order phase transitions in the annealing path to solve the optimization problem 
efficiently using QA algorithm. Below we discuss various methods to speed up a quantum annealing process.

\subsection{Application of antiferromagnetic fluctuations}
\label{anti_fluctuations}
In the context of speed-up of QA, Seki and Nishimori in 2012
proposed a method~\cite{seki2012quantum} to overcome issues related to first order phase transitions, 
by studying quantum annealing in {the} presence of antiferromagnetic fluctuations in 
addition to the transverse-field term. They applied the 
method to the infinite-range ferromagnetic $p$-spin model (see Eq.~(\ref{p-spin model}))
and showed that there exists a quantum path that avoids 
first-order transitions for some intermediate values of $p$. 
The Hamiltonian for $p$-spin is given by
\begin{equation}
 H_{0} = -N\Bigl(\frac{1}{N}\sum _{i=1}^{N} \sigma _{i}^{z}\Bigr) ^{p}.\label{eq:p-spin model}
\end{equation}
This is indeed the classical counterpart of the Hamiltonian as in Eq.~(\ref{p-spin model}) with zero 
transverse field.
Here $H_{0}$ is the 
target Hamiltonian $H_P$,  whose ground state is the optimal solution 
of the problem. The QA for this model is studied before with
the transverse-field as a driver Hamiltonian $H_D$, which takes 
an exponentially long time to reach the ground state of the 
target Hamiltonian due to the presence of a quantum first-order 
phase transition during the time evolution. The ferromagnetic 
$p$-spin model reduces to the Grover problem when
$p\rightarrow\infty$ and there is no known algorithm that can 
solve the problem efficiently {in a polynomial time}.

We discuss here how {the inclusion of} an antiferromagnetic fluctuation term 
can improve the performance of QA of the model {when both
the transverse-field term and the antiferromagnetic term are tuned.} The total Hamiltonian of 
the problem is then given by
\begin{equation}
 H (s,\lambda )
  = s\{ \lambda H_{0} + (1-\lambda )\hat{V}_{\!\text{AF}} \}
  + (1-s)V_{\text{TF}},\label{ham_s-lambda}
\end{equation}
where $V_{\!\text{AF}} =
  +N \Bigl(\frac{1}{N}\sum _{i=1}^{N} \sigma _{i}^{x}\Bigr)^{2}$ is an antiferromagnetic interaction term, whereas 
 $V_{\rm TF}$ is the conventional transverse-field term. 
The parameters $s$ and $\lambda$ are functions of time, { assumed to lie between $0$ and $1$}, which 
are chosen appropriately for a QA process. 
The initial Hamiltonian 
is defined by $s=0$ and an arbitrary $\lambda$, and the final one 
is given by $s=\lambda=1$. 

% \subsection{Static approximation and low-temperature limit}
% \label{low_temp}
% We now focus on possible analytic computations of the system and 
% find out phase diagrams on the $s-\lambda$ plane for different 
% values of $p$. 
\subsubsection{Numerical results}
\label{num_results}
\begin{figure}[t]
  \centering
  \includegraphics[width=6cm, bb = 0 0 226 141]{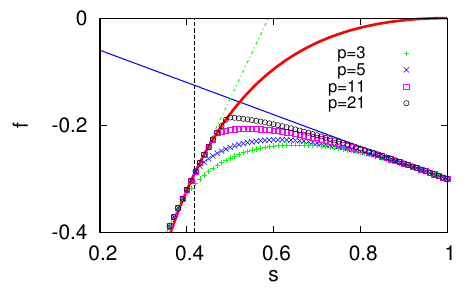}
  \includegraphics[width=6cm, bb = 0 0 226 141]{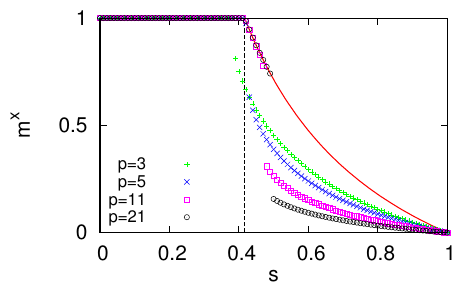}\\
\caption{\label{fig:free energies}
  Left panel: Free energy of the system with the Hamiltonian in 
  Eq.~(\ref{ham_s-lambda}), as a function of $s$ for some values 
  of $p$ with $\lambda=0.3$. The free energy of the QP phase, Eq.~(\ref{f_QP_static}), 
  is represented by the dash-dotted line in light green. The thin solid line with blue color represents the 
  free energy of the ferromagnetic phase (F), Eq.~(\ref{f_F_static}), and the thick solid line with red color 
  is for the QP2 phase, Eq.~(\ref{f_QP2_static}). The lower limit 
  of the QP2 domain ($s=1/(3-2\lambda )$) is denoted by the vertical dashed line. Although it is hard to 
  see in this present scale, all the data for finite $p$ studied here, have lower values than that 
of $f_{\text{QP2}}$ in the QP2 regime.
Right panel: Magnetization $m^{x}$ vs. $s$ for $\lambda=0.3$ and the same values of $p$ as in the case of 
free energies. The verical dashed line is the same as shown in the left panel. The solid line in red exhibits 
the $x$~component of magnetization of the QP2 phase. The magnetization decreases to zero as $s$ is increased 
with a jump at the boundary of the QP2 domain for $p \ge 5$.
Taken from~\cite{seki2012quantum}.
  }
 \end{figure}

We now focus {on} analyzing the phase diagram of the model on the 
$s-\lambda$ plane for finite values of $p$, using {the} saddle point 
method and {the} static approximation, {as elaborated in Appendix~\ref{stat_approx_appendix}}. The method we adopt
to construct the phase diagram as follows. The self-consistent 
equations (\ref{self_consistent_eq_mz_ltemp}) and (\ref{self_consistent_eq_mx_ltemp}) 
%{(see Appendix~\ref{stat_approx_appendix})} 
are initially solved numerically for a particular value of $p$
and a set of values of $s$ and $\lambda$ to find out corresponding 
free energy. By comparing these free energies and all possible 
solutions, the stable phases of the system are identified with 
smallest value of free energy. The variation of free energy with 
$s$ for some values of $p$ and $\lambda=0.3$ is shown in Fig.~\ref{fig:free energies}.
It can be seen that the free energies for different values of $p$ 
lie below $f_{\rm QP2}$ in the ${\rm QP2}$ regime and, therefore, 
the ${\rm QP2}$ phase is not a stable phase. As we vary $\lambda$, 
the system undergoes a quantum phase transition from the QP phase 
for small $s$ to the  {Ferromagnetic (F)} phase for large $s$.

%%%%%%%%%% free energy     %%%%%%%%%%%%%%%%%
%%%%%%%%%%%%%%%%%%%%%%%%%%%%%%%%%%%%%%%%%%%%

To determine the type of phase transitions, i.e., first or second 
order, the magnetization $m^x$ is numerically calculated as a 
function of $s$ for the same parameter values as in the case of 
free energy. We observe a change in $m^x$ around $s=0.4167$ and 
$m^x$ decreases continuously to zero from its unit value for $p\ge5$. 
Equivalently, it indicates that $m^z$ increases continuously 
from zero to a finite value as we increase $s$ for $p\ge5$. 
This identifies that for $p\ge5$ there exists a second order phase 
transition at the boundary of ${\rm QP}$ and ${\rm QP2}$ phases.
An interesting scenario arises for some parameter values 
(e.g., $\lambda=0.3$, $p=11$) where the magnetization shows a jump 
within the ferromagnetic phase. This suggests the existence of 
first-order transition within the F phase and the energy gap at 
the transition point decreases exponentially with the system size. 
Nevertheless, this peculiar behavior does not appear for smaller 
values of $\lambda$ for any non-zero $p$, except $p=3$. 
Therefore, for smaller $\lambda$, there exists only a second 
order transition when one increase $s$ from zero to a value 
near unity. 

\begin{figure*}[t]
%  \centering
\begin{center}
 \includegraphics[width=\hsize, bb = 0 0 453 147]{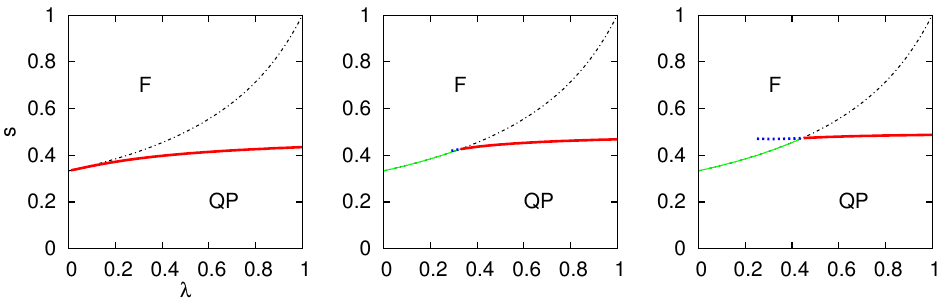}
\end{center}
  \caption{\label{fig:phase diagram}%
Phase diagrams on the $s$-$\lambda$ plane for the system with Hamiltonian 
in Eq.~(\ref{ham_s-lambda}) for $p=3$ (left), $p=5$ (middle), and $p=11$ (right). The boundary of the QP2 domain 
($s=1/(3-2\lambda )$), where a transition occurs between the 
QP and F phases, is represented by the dash-dotted line. 
The red lines are for first-order transitions
and the light green lines represent second-order transitions.
For $p=5$ and $11$, the magnetization shows sudden jumps on the 
dashed blue line (F-F boundary) within the F phase.
Taken from~\cite{seki2012quantum}.}
 \end{figure*}

Using these results, phase diagrams of the system for $p=3, 5,$ 
and $11$ are drawn on the $s-\lambda$ plane (see Fig.~\ref{fig:phase diagram}). 
We can see that a boundary of second-order transitions between 
F and QP phases exists for small $\lambda$ and $p\ge5$. As a 
consequence, there are possibilities to find a path to reach the F phase 
from the QP phase by avoiding a first-order transition provided 
the first-order F-F boundary does not reach the $\lambda=0$ 
axis, that occurs probably in the limit of $p\rightarrow\infty$~\cite{seki2012quantum}.

\begin{figure}[t]
%  \centering
\begin{center}
  \includegraphics[width=6cm, bb = 0 0 226 141]{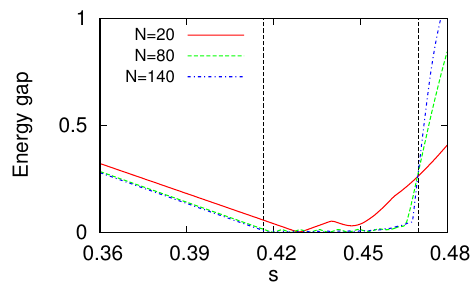}
  \includegraphics[width=6cm, bb = 0 0 226 136]{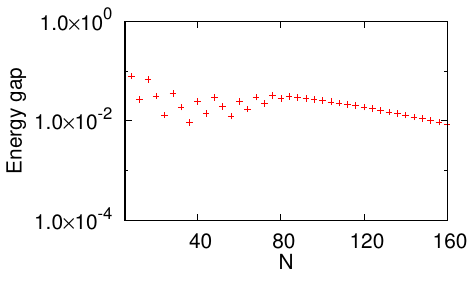}
\end{center}
  \caption{\label{enegy_gap_lam=0.3}%
  Left panel: Energy gap as a function of $s$ for $p=11$ and $\lambda =0.3$. The positions of minima of the energy gap are 
  shown by the vertical dashed lines at the QP2 domain
  with $s\simeq 0.4167$ and the F-F boundary at $s\simeq 0.4701$.
  Right panel: The rightmost local minimum of the energy gap with 
  the system size $N$ for $p=11$ and $\lambda =0.3$ on a semi-log scale. 
  The energy gap closes exponentially fast with $N$. Taken from~\cite{seki2012quantum}.}
 \end{figure}

Let us now focus on analyzing the behavior of the energy gap 
across the phase transition points of the system. The energy 
gap of the system is calculated numerically using perturbation 
theory as described in Ref.~\cite{jorg2010energy}. The variation 
of energy gap with $s$ for $\lambda=0.3$ and $p=11$ is shown in 
Fig.~\ref{enegy_gap_lam=0.3}. If the range of $s$ where the energy 
gap has minimum value is {zoomed}, it can be seen that the gap 
shows wiggly behavior throughout the range. This behavior starts 
at $s\simeq 0.4184$ for $\lambda=0.3$, which indeed corresponds {to}
the second-order transition point between the QP and F phases. 
The wiggly behavior ends at $s\simeq 0.4676$ for $\lambda=0.3$, 
which corresponds to the first-order transition point at the 
F-F boundary. The dashed vertical lines in Fig.~\ref{enegy_gap_lam=0.3} 
indicate two transition points that are evaluated analytically 
using Eqs.~\ref{self_consistent_eq_mz_ltemp} and \ref{self_consistent_eq_mx_ltemp}. 
The analytical results show nearly a good agreement with the numerical 
results in the interval where the gap is very small. 
It has been found that the rightmost local minimum of the energy gap in Fig.~\ref{enegy_gap_lam=0.3} 
corresponding the F-F boundary shows different scaling relation with 
the system size $N$ compared to the other local minima. The rightmost 
minimum energy gap decays exponentially with the system size, which 
is expected from discontinuous behavior of the magnetization in 
Fig.~\ref{fig:free energies} at the F-F boundary implying the 
first-order transition. Although, for the present case, the above 
mentioned energy gap is not a global minimum, this will affect the 
efficiency of QA for larger systems where the rightmost gap can 
become a global minimum since the other local minima decay 
ploynomially with the system sizes (see Figs. $6$ and $7$ of 
\cite{seki2012quantum} for details).

These analytical and numerical results suggest that it is 
possible to increase the efficiency of QA by choosing a path 
around $\lambda=0.1$, which avoids first-order transition to 
reach the F phase from the QP phase. For this process, $s$ is 
the tuning parameter and the value $p$ needs to be  chosen  within 
the range $5\le p\le21$ achieving  maximum efficiency.

%%%%%%%%%%  phase diagrams %%%%%%%%%%%%%%%%%
%%%%%%%%%%%%%%%%%%%%%%%%%%%%%%%%%%%%%%%%%%%%

%%%%%%%  energy gap for p=11,lambda=0.3  %%%
%%%%%%%%%%%%%%%%%%%%%%%%%%%%%%%%%%%%%%%%%%%%

%SSSSSSSSSSSSSSSSSSSSSSSSSSSSSSSSSSSSSSSSSSSSSSSSSSSSSSSSSSSSSSSSS
\subsection{Inhomogenous transverse field driving}
\label{inhomogenous_TF}
In Sec.~\ref{sp_qa}\ref{anti_fluctuations}, we have discussed a method 
to speed up a QA process in {the} presence of first order phase transitions by adding an antiferromagnetic fluctuation term. 
Here we consider {a relatively} simpler approach of inhomogenous 
driving of the transverse field to overcome the issue of 
first order phase transitions. In this case, the strength of 
transverse field is turned off sequentially from one site 
to the next according to the annealing schedule. Using both 
analytical and numerical calculations, it has been shown that 
inhomogenous driving can completely remove QPTs from the 
annealing path and thus, it speeds up the annealing process 
exponentially~\cite{susa2018exponential,susa2018quantum}. 

The total Hamiltonian for the inhomogenous driving is given 
by
\begin{equation}
    H(s, \tau)=s H_0-\sum_{i=1}^{N(1-\tau)}\sigma_i^x,
    \label{inho_driving_Ham}
\end{equation}
where $H_0$ is the Hamiltonian for $p$-spin model 
in Eq.~(\ref{eq:p-spin model}). The parameters $s$ and $\tau$ 
both are time-dependent, where $s=\tau=0$ at $t=0$ and 
$s=\tau=1$ at $t=t_0$. This shows that the initial Hamiltonian 
has only transverse field and the final one has only $p$-spin 
interacting term with the Hamiltonian $H_0$. Both the initial 
and final Hamiltonians are in agreement with the traditional 
QA protocol.

We note that the transverse field in Eq.~(\ref{inho_driving_Ham}) 
is applied only to $N(1-\tau)$ spins, where $\tau$ increases from 
$0$ to $1$ as time proceeds from $0$ to $t_0$. This indicates that 
the transverse field is turned off {at neighbouring sites one by one} as time increases, 
starting from site $i=N$ to ending with site $i=1$ at $\tau=1$. 
This is the process of how the transverse field is driven inhomogeneously. 
It can be noted that the parameter $\tau$ can 
have only discrete values for a finite $N$, since the upper 
limit $N(1-\tau)$ in Eq.~(\ref{inho_driving_Ham}) should be an 
integer.

\subsubsection{Results}
\label{inhomogenous_TF_results}
\begin{figure}[t]
%\centering
\begin{center}
\includegraphics[width=0.7\linewidth, bb = 0 0 381 272]{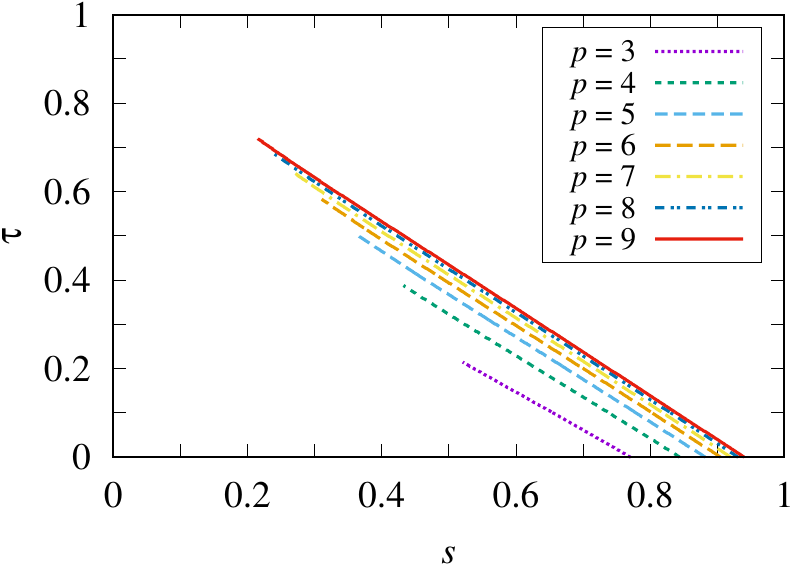}
\end{center}
\caption{Phase diagram of the $p$-spin ferromagnetic model under inhomogenous field. 
All the lines represent first-order phase 
transitions for different values of $p$, which are extended up to 
the middle of the phase diagram from the axis $\tau=0$ corresponding 
the homogenous model. Taken from~\cite{susa2018exponential}.}
\label{fig:phase_diagram_inhomo}
\end{figure}

Using Trotter decomposition and the static approximation in 
Hamiltonian (\ref{inho_driving_Ham}), the free energy of the 
system can be calculated analytically for both finite and zero 
temperatures (see appendix~\ref{inho_driving_appendix}). 
By minimizing the 
zero-temperature free energy with respect to magnetization $m$ 
produces a ground state phase diagram as depicted 
in Fig.~\ref{fig:phase_diagram_inhomo}.

For a fixed value of $p$, a line of first-order phase transitions 
originated from a point on the $s$-axis, terminates before approaching to any point 
of the $\tau$-axis. Remarkably, all these 
lines for different  values of $p$ end before they reach one of the axes, 
$\tau=1$ or $s=0$. Therefore, there exists a path starting from 
$s=\tau=0$ to $s=\tau=1$, that does not encounter any kind of 
phase transitions. This leads to an exponential speedup of QA, 
since the energy gap always remains finite even for a system with 
large size. The positions of the critical points on the $\tau-s$ 
plane where the first-order transitions terminate for different 
$p$ values, are calculated analytically by using the standard 
Landau theory of phase transitions (see Eq.~(\ref{location_first_transitions})). 
In this calculation, it has 
been considered that the coefficients of the expansion of the free energy (\ref{f_T0_inhomo}) 
around its minimum at $m=m_{\rm c}$ vanish to third order~\cite{nishimori2010book}.

\begin{figure*}
%  \centering
\begin{center}
%  \subfigure[\ ]{\includegraphics[width=0.32\linewidth]{figs/energy_gap_p_3_r_1_inhomo.eps}\label{fig:energy_gap-a_inhomo}} 
 % \subfigure[\ ]{\includegraphics[width=0.32\linewidth]{figs/energy_gap_p_3_r_2_366_inhomo.eps}\label{fig:energy_gap-b_inhomo}} 
  %\subfigure[\ ]{\includegraphics[width=0.32\linewidth]{figs/energy_gap_inhomo.eps}\label{fig:energy_gap-c_inhomo}}
%   \includegraphics[width=0.32\linewidth]{figs/energy_gap_p_3_r_1_inhomo.eps}\label{fig:energy_gap-a_inhomo}
%  \includegraphics[width=0.32\linewidth]{figs/energy_gap_p_3_r_2_366_inhomo.eps}\label{fig:energy_gap-b_inhomo} 
%  \includegraphics[width=0.32\linewidth]{figs/energy_gap_inhomo.eps}
%\label{fig:energy_gap-c_inhomo}
   \includegraphics[width=0.32\linewidth, bb = 0 0 306 264]{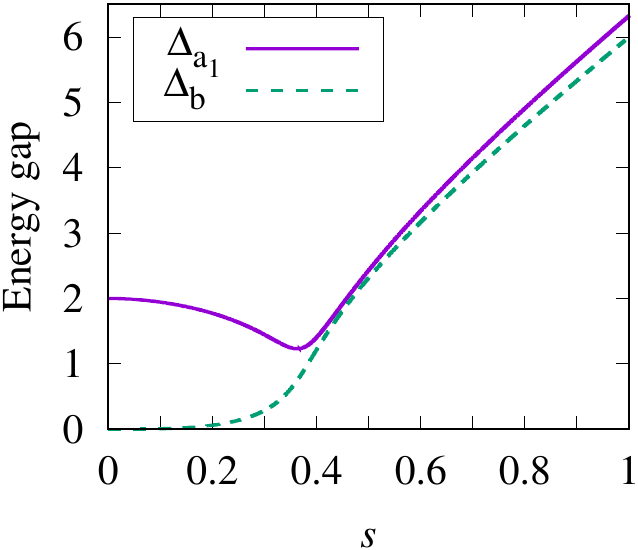}\label{fig:energy_gap-a_inhomo}
  \includegraphics[width=0.32\linewidth, bb = 0 0 306 264]{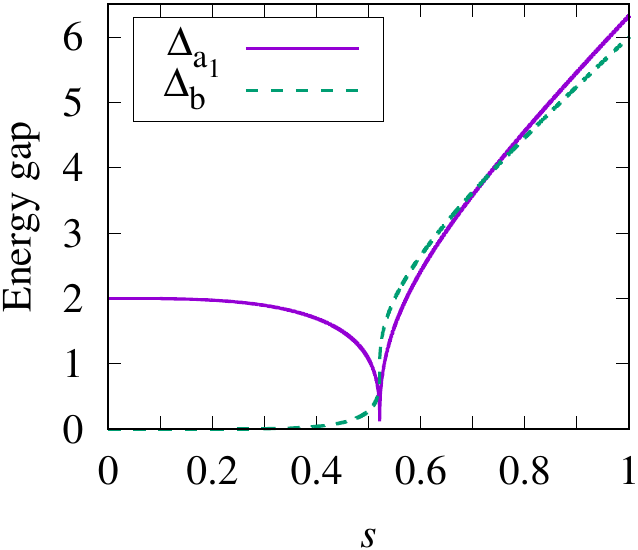}\label{fig:energy_gap-b_inhomo} 
  \includegraphics[width=0.32\linewidth, bb = 0 0 306 264]{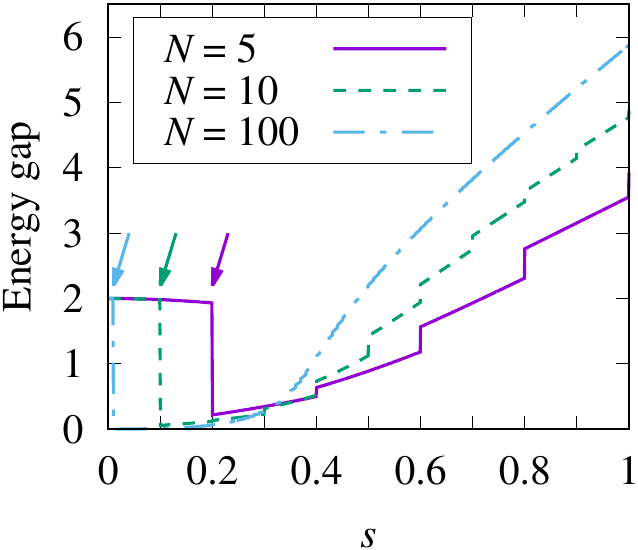}
\end{center}
\label{fig:energy_gap-c_inhomo}
  \caption{Plots of two types of energy gap $\Delta_{a_1}$ and $\Delta_{b}$ for $p=3$ 
  as functions of $s$ with (left) $\tau = s$ (i.e., away from the transition line) and (middle) $\tau = s^{2.366}$ 
  (i.e., just touches the critical point). The final energy gap is defined by the smaller of these two gaps. 
  Right: The energy gap for different system sizes with $\tau=s$ computed by direct numerical diagonalization. 
  The location of the minimum energy gap is shown by an arrow for each $N$. Taken from~\cite{susa2018quantum}.}
   \label{fig:energy_gap}
\end{figure*}

To strengthen the above conclusion, the energy gap of the system has 
been calculated both analytically and numerically. Since our system 
is mean-field-type, the semi-classical treatment can be applied
to evaluate the energy gap~\cite{seoane2012many,filippone2011quantum}.
In this context, the parameterization of a path $\tau=s^r$ is considered to connect $s=\tau=0$ and $s=\tau=1$ with a parameter 
$r$ that determines the shape of the path. Figures~\ref{fig:energy_gap-a_inhomo} and \ref{fig:energy_gap-b_inhomo} 
exhibit two energy gap candidates, $\Delta_{a_1}$ and $\Delta_{b}$, for the system with $p=3$ along the paths $\tau=s$, 
that does not encounter phase transitions, and $\tau = s^{2.366}$, that just touches the critical point where the first-order line ends. 
The smaller one between these two candidates is the actual 
energy gap of the system.

As shown in Fig.~\ref{fig:energy_gap-a_inhomo}, $\Delta_{b}$ 
is found to be smaller one and, it monotonically increases 
with $s$. On the other hand, as expected, the energy gap 
$\Delta_{a_1}$ vanishes at the critical point $s_c\approx0.52$. 
To investigate the effect of finite-size systems, the energy 
gap is calculated by a direct numerical diagonalization method 
along the $\tau =s$ path. The result is shown in 
Fig.~\ref{fig:energy_gap-c_inhomo}, which shows a very good 
agreement with the asymptotic behavior in the $N\to\infty$ limit 
as observed in in $\Delta_b$ of Fig.~\ref{fig:energy_gap-a_inhomo}. 
As seen in Fig.~(\ref{fig:energy_gap-c_inhomo}), the energy gap 
becomes minimum when the transverse field is turned off at the 
first site as shown by the arrows, thus implying the location 
of the minimum gap at $s=0$ in the $N\to\infty$ limit. 
It is important to note that the the energy gap becomes minimum 
near the origin $\tau=s=0$, when the annealing path ($\tau=s$) 
does not encounter any transitions (see Figs.~\ref{fig:energy_gap-b_inhomo} 
and \ref{fig:energy_gap-c_inhomo} ), whereas the minimum of the 
gap occurs at the critical point when such a transition exists 
along the path ( see Fig.~\ref{fig:energy_gap-b_inhomo}).
In addition, a series of paths is considered to examine the 
inhomogenous driving protocol, and it is found that the minimum 
energy gap shows an exponential decrement with the system sizes 
when the paths cross the first order transitions 
(for details, { refer to the} discussion around Eq.~($12$) of 
Ref.~\cite{susa2018quantum}). 

The problem that we discussed so far, for inhomogeneous driving,
considers ideal situations, i.e., zero temperature and a complete
turning off of the transverse field at each site. This problem 
also has been studied at finite temperature and zero temperature 
with different types of inhomogenity. It has already been studied 
that the first-order transitions that exists under homogeneous 
driving at zero temperature can be circumvented by inhomogeneous 
driving with complete turning off the transverse field at each 
site. For nonideal situations, with a finite temperature or a 
non-zero value of the final transverse field, one can not avoid 
new first-order transition lines like the ideal case. 
Nevertheless, it has been observed that the new first-order 
transitions are weaker than the original one, since the free 
energy barrier between two arbitrary local minima is smaller 
than the original homogeneous case. This leads  to an increase 
in the quantum tunneling rate. Therefore one can infer that 
the inhomogeneous driving of the transverse field has the 
potential to provide a better performance in quantum annealing.
In addition, Matsuura et al.~\cite{matsuura2016mean} 
studied analytically  the p-body ferromagnetic infinite-range Ising
model in transverse-field using a mean-field analysis and demonstrated
that for $p \ge 3$, where the phase transition is of first order, 
Quantum Annealing Correction softens the closing of the gap for small 
energy penalty values and prevents its closure for sufficiently
large energy penalty values, thereby providing  from excitations that
occur near the quantum critical point. It also has been shown analytically
that nested quantum annealing correction can suppress errors effectively 
in  Ising models with infinite-range interactions and their analysis revealed 
that the nesting structure can significantly weaken or remove the
first-order phase transitions, where the energy gap closes exponentially~\cite{matsuura2019nested}.

\subsection{Suppression of Griffiths singularities}
\label{supp_gr_singu}
In a recent work by Knysh et al.~\cite{knysh2020quantum}, it has 
been shown that a QA process can be accelerated using an embedded 
spin chain system with random interactions. A randomly interacting 
spin chain exhibits Griffiths-McCoy singularities~\cite{griffiths1969nonanalytic,mccoy1969incompleteness}, 
since different parts of the system can not reach criticality simultaneously for 
random fluctuations. This leads to the diverging dynamical exponent 
$z$ and a stretched exponential scaling of the energy gap~\cite{fisher_random_1992}. 
Therefore the presence of Griffiths singularities increases the annealing time 
for such systems. 

On the other hand, quantum annealing has been studied for a embedded spin 
chain problem, where logical qubits were replaced by ferromagnetic Ising 
spin chains~\cite{knysh2020quantum}. For this study, an ansatz is considered to find a balanced 
choice of coupling parameters based on renormalization group intuition 
for the better performance of QA. This results an exponential improvement 
of annealing time, which is also confirmed numerically. It indicates 
that this protocol prevents to occur randomly oriented domains in the 
system by ensuring a simultaneous criticality of spatially separated regions.

\section{Application}
\label{applications}
In this section, we discuss 
some recent results on QA of a 
initeracting bosonic system coupled with cavity field modes for
both classical and quantum limits of the system.
In addition, we will mention about the recent development of 
QA in context of parallel computation, and its effectiveness 
compared to the existing methods.

\subsection{Quantum annealing vs. semi-classical annealing}
\label{QA_CA}
%\tr{ We have to compare with what Sei has written}. The idea of QA 
%that was originally proposed for spin glass systems by Ray et al., 
%recently, has been applied for interacting bosonic systems coupled 
%to cavity modes by Starchl et al.~\cite{starchl2022unraveling}.  
%In "Introduction" section of this paper, the authors quoted about 
%the $1989$ work by Ray et al. on quantum tunneling in the spin 
%glass phase, as, {\it Notably, one of the earliest work in laying 
%the foundation of quantum annealing, was done in 1989 [6],showing 
%that quantum fluctuations can increase the ergodicity in a spin-
%glass model, by tunneling between ‘trapping’ minima, sepa-
%rated by narrow potential barriers.} 
Starchl and Ritsch~\cite{starchl_unraveling_2022},  have used the 
idea of quantum tunneling to show the success of QA over 
semi-classical annealing for an interacting bosonic model in 
presence of cavity modes. The model that is considered for this study is described by a 
tight binding Bose-Hubbard lattice model with four sites, which 
are filled by two interacting bosons. The tunable non-local 
interactions are introduced in the model via collective light 
scattering to two independent cavity modes. The Hamiltonian 
for such system is given by 

\begin{eqnarray}
	H & = J\sum_{k_{PBC}}(b_k^\dagger b_{k+1} + h.c.)  +\frac{U}{2} \sum_k n_k(n_k - 1)  \nonumber  \\
         -&\Delta (a_1^\dagger a_1 + a_2^\dagger a_2) +\tilde{J}\big(\hat{M}_1 (a_1+a_1^\dagger) 
     + \hat{M}_2 (a_2+a_2^\dagger)\big), \label{eq:Bose-Hubbard}
\end{eqnarray}
where $b_k$ and $b_k^{\dagger}$ are bosonic annihilation and 
creation operators, respectively. The photonic annihilation operators $a_1$ and $a_2$ are associated with two independent 
cavity modes. The interactions 
between the bosons and cavity field modes are represented by the 
fourth term of Eq.~(\ref{eq:Bose-Hubbard}), where $\hat{M_1}$ and 
$\hat{M_2}$ are called effective scattering operators. Using 
mean-field approximation of the field operators, the Hamiltonian 
in Eq.~(\ref{eq:Bose-Hubbard}) can be written in semiclassical 
form
\begin{equation}
	H^{sc}= \frac{\Delta \tilde{J}^2}{\kappa^2 + \Delta^2}\big( 2\hat{M}_1\langle\hat{M}_1\rangle - \hat{{I}}\langle\hat{M}_1\rangle^2 +   2\hat{M}_2\langle\hat{M}_2\rangle - \hat{{I}}\langle\hat{M}_2\rangle^2 \big),
	\label{eq:Ham-semi-class}
\end{equation}
where $\kappa$ determines the strength of cavity loss, where $I$ denotes the identity operator. Both the 
Hamiltonians in Eqs.~(\ref{eq:Bose-Hubbard}) and (\ref{eq:Ham-semi-class}) with periodic boundary conditions are translationally 
invariant and thus provide approximately degenerate ground states. 
In order to create an unique target ground state for the annealing 
process, a certain amount of impurity of strength $V$ is added 
in the Hamiltonian. 

\begin{figure}[h!]
\centering
\includegraphics[width=0.47\textwidth, bb = 0 0 480 360]{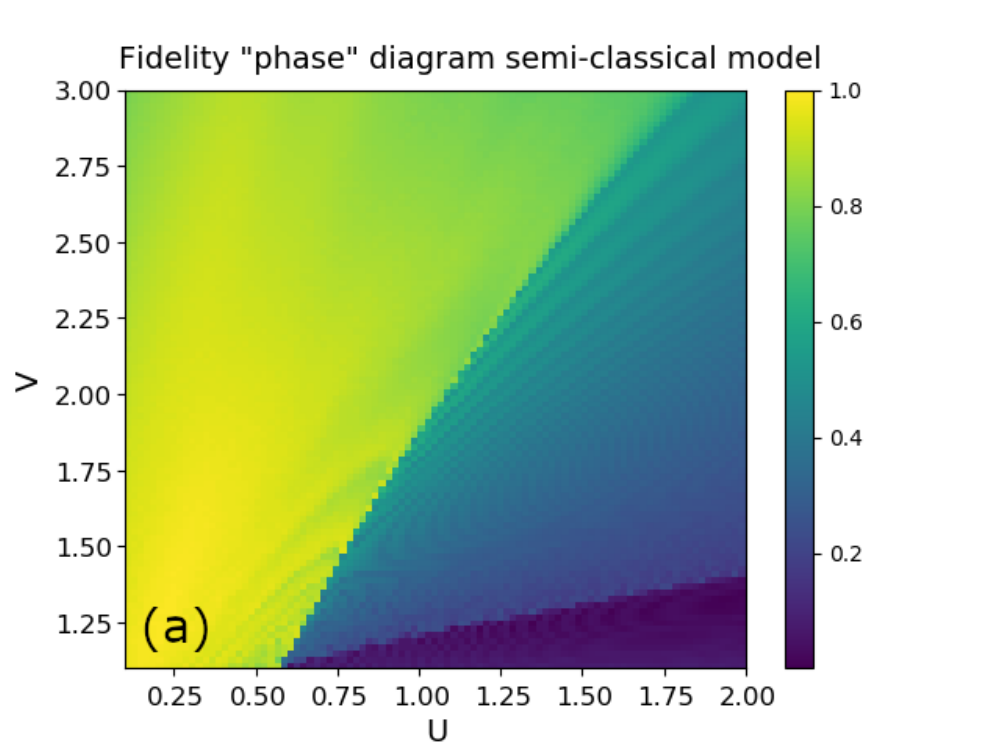} 
\includegraphics[width=0.47\textwidth, bb = 0 0 480 360]{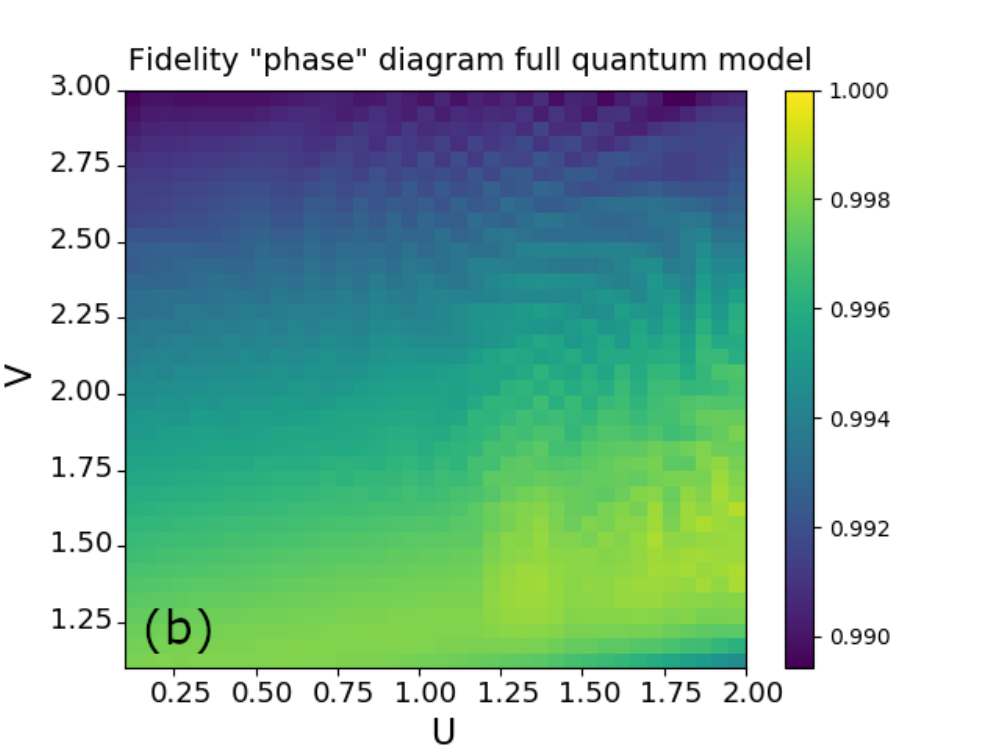}
\caption{Color density plot of the fidelity calculated as the overlap between the final state 
after an adiabatic evolution using (a) the Hamiltonian with semi-classical mean-field 
approximation and as well as (b) for the full quantum Hamiltonian, and the desired target 
state on the $U-V$ plane. Here the parameter values are: $t_f=1000$. For (a): $\Delta = -1$, $\tilde{J}=1$ and 
for (b): $\Delta = -5$, $\tilde{J}=\sqrt{5}$. Taken from~\cite{starchl_unraveling_2022}.}
\label{fig:fidelity_annealing}
\end{figure}

The dynamics of the system is started with the ground state at 
zero pump $\tilde{J} = 0$, and the pump strength is increased 
linearly towards $\tilde{J} = \sqrt{5}$, following an adiabatic schedule: 
$\tilde {J} \approx t/t_f$, where $t_f$ is the final time. The results of the study of annealing for this system 
are summarized in Fig.~\ref{fig:fidelity_annealing}. It shows 
the density plot of the fidelity on $U-V$ plane after an adiabatic sweep using both the semi-classical (\ref{eq:Bose-Hubbard}) 
and the full quantum (\ref{eq:Ham-semi-class}) Hamiltonian. 
In this case, the fidelity is calculated as the overlap between 
the final state after an adiabatic sweep and the desired target state. By comparing two density plots for both semi-classical 
and quantum cases, i.e., when the dynamics is driven by the 
semi-classical Hamiltonian and the full quantum Hamiltonian, 
respectively,
%\tr{it is not clear what is semi-classical and quantum; please elaborate; what I understand that study the dynamics of full quantum Hamiltonian versus the semi-classical form of the Hamiltonian}, 
one can identify a clear quantum improvement 
in the success rate. It is observed that the semi-classical 
approximation provides a reliably correct solution for small 
onsite interaction strength. For this scenario, one can see 
a sudden fall of the success rate, when the interaction strength 
is increased. If the repulsive interaction is further increased, 
the gap between the final ground state energy and the first 
excitation becomes very small and the classical model effectively 
never succeeds. On the other hand, the adiabatic evolution with 
full quantum Hamiltonian (\ref{eq:Bose-Hubbard}) provides almost 
correct solution with $99\%$ probability, even for higher $U$ 
values as shown in Fig.~\ref{fig:fidelity_annealing}(b). 
Therefore, for this system, a large parameter region is found 
where quantum annealing is highly successful, whereas the 
semi-classical approach largely fails. In addition, for quantum 
scenario, a direct connection is found between atom-field 
entanglement in the dynamics and a high probability to find 
the correct solution at end of annealing process.

\subsection{Parallel quantum annealing}
\label{parallel_QA}
There are a few recent studies on QA in the context of parallel 
quantum computation~\cite{jalowiecki2020parallel,aadit2022massively,pelofske2022parallel}.
Recently, Pelofske et al.~\cite{pelofske2022parallel} propose 
a method named as parallel quantum annealing that has potential 
to solve many independent problems on a quantum annealer during 
the same annealing process. The authors applied their proposed 
method of parallel quantum annealing on both D-Wave $2000Q$ 
at Los Alamos National Laboratory (refereed to as D-W $2000Q$) 
and the newer D-Wave ${\rm Advantage\textunderscore System}$ $1.1$ (referred to as D-W Advantage). 
The results of parallel quantum annealing have been 
compared with those found from sequential quantum annealing, i.e., 
when the same problems are solved sequentially on D-Wave machine. 
It has been observed that, although, there is a slight decrement 
in the accuracy of the solution for simultaneously solved problems, 
parallel quantum annealing can provide a considerable speedup 
of up to two orders of magnitude~\cite{pelofske2022parallel}.

\section{Summary and outlook}

We have provided an overview of  recent developments of QA which is based on the possible advantage
of utilizing quantum tunneling. When the energy landscape of an Ising 
Hamiltonian, where the corresponding ground state is the target state of an optimisation problem, consists of high but thin barriers surrounding local minima, 
quantum tunneling has an advantage over thermal fluctuation in overcoming 
barriers and thus getting the system equilibrated. This nature of quantum 
tunneling provides a  foundation which asserts that  QA can outperform SA in a glassy system 
with a rugged energy landscape. Indeed, we have  focused on  analytical and numerical evidences 
that QA yields a better solution than SA in several glassy systems.
 %probably due to the advantage of 
%quantum tunneling in a glassy system.
 The question of the restoration of 
ergodicity due to quantum tunneling in the quantum SK model is still unresolved. 
Nevertheless, recent studies do indicate the possibility of  the existence of an ergodic phase 
at least in low temperature region~\cite{mukherjee2015classical,Leschke_2021}: this
is expected to lead to a remarkable possibility of the success of the  annealing scheme in those systems~\cite{mukherjee2018possible}. 

We have  also discussed quantum phase transitions (QPTs) in connection with QA. 
Generally speaking, a QPT hinders the adiabatic time evolution underlying QA, 
since the energy gap above the ground state closes at a QPT; this leads to inevitable generation of defects and excess energy in the final evolved state. In this context, it is worth noting that 
the short-cut to adiabaticity \cite{guery-odelin_shortcuts_2019} or counter-diabatic driving protocols \cite{Sels_2017}  have been proposed as 
a method to realize the adiabatic time evolution with a finite time,
providing  a possible route to avoid a continuous QPT in the process of annealing. On the contrary,
a discontinuous QPT involves an exponentially fast closure of the energy gap 
with  the system size. Therefore, it is desirable and at the same time far more challenging  to circumvent 
a discontinuous QPT for the success of QA. 
We have  reviewed methods using additional 
antiferromagnetic interactions~\cite{knysh2020quantum} or inhomogeneous 
transverse fields~\cite{susa2018exponential,susa2018quantum} proposed for  avoiding
discontinuous QPTs which would lead to  acceleration of QA. 
Random interactions in spin chain systems induce Griffiths singularities that 
eventually increase annealing time for such systems. These singularities can be 
suppressed for the  embedded spin chain problem, where ferromagnetic Ising spin chains 
are used as logical qubits. It has been shown that the performance of QA for such embedded
systems improves exponentially in the context of annealing time~\cite{knysh2020quantum}.
To be precise, the performance advantage of QA is
still model specific and a generic prescription has not been known for discontinuous phase transitions
involved by practical optimization problems. In the cases of spin glass models, however,
major advantages of the standard QA have now even been established using QMC simulation \cite{santoro_theory_2002} and hardware implementations \cite{brooke_quantum_1999,king_quantum_2022,mohseni2022ising}.

The idea of quantum tunneling has been further used by Starchl and Ritsch~\cite{starchl_unraveling_2022}  to 
establish  the superiority  of QA over semi-classical annealing for a realistic system of interacting bosons 
in presence of cavity modes. Using numerical results, it has been shown that there exists a large parameter 
regime where the QA provides a better performance than semi-classical annealing. In addition, we have 
noted the recent development of parallel quantum computation using an annealing algorithm~\cite{pelofske2022parallel}. 
The idea of parallel QA is to solve many independent problems on a quantum annealer during the same 
annealing schedule. The authors have checked their method of parallel QA on D-wave quantum machines 
and indicated the effectiveness of the same \cite{pelofske2022parallel}.

Finally, in recent years  the progress in developing hardware that performs QA using physical qubits has gained a tremendous momentum. 
So far, devices with more than 5000 qubits have been made available, and employed to study a wide gamut of fields that include
condensed matter systems in and out of equilibrium \cite{harris_phase_2018,kairys_simulating_2020,bando_probing_2020,king_quantum_2021,king_quantum_2022}, high-energy physics \cite{mott_solving_2017,das_track_2019,abel_2021}, quantum chemistry \cite{teplukhin_calculation_2019}, 
and biology \cite{perdomo-ortiz_finding_2012,li_quantum_2018}. Application to various optimization 
problems has been developing as well. We reviewed some of the experimental studies using a QA hardware. 
From the viewpoint of application as well as gaining theoretical rigor, decoherence inherent in a device coupled to an environment   is a fundamental  issue of interest. 
We have briefly reviewed effects of thermal and non-thermal environments on QA. In order to perform QA ideally, 
coupling to an environment leading to decoherence should be reduced. However, environment assisted QAs 
have been proposed for specific situations \cite{amin_thermally_2008,dickson_thermally_2013,mishra_finite_2018}. 
Utilising specially engineered environments to accelerate QA would be an important direction of future study.

%\enlargethispage{20pt}

\appendix

\numberwithin{equation}{section}
\section{Static approximation and low-temperature limit}
\label{stat_approx_appendix}
Using Suzuki-Trotter formula, the partition function for the 
Hamiltonian in Eq.~(\ref{ham_s-lambda}) can be written as
\begin{align}
 Z &= \lim_{M\to \infty} Z_{M}
\notag \\
&=  \lim_{M\to \infty }{\rm Tr} \bigl(e^{-\frac{\beta}{M} s\lambda H_{0}}
  e^{-\frac{\beta}{M}\{ s(1-\lambda )V_{\!\text{AFF}}
  + (1-s)V_{\text{TF}}\}}\bigr)^{M}
\notag \\
&= \lim_{M\to \infty }\sum _{\{ \sigma _{i}^{z}\}} \langle \{\sigma _{i}^{z} \}|
\biggl(\exp \Bigl[\frac{\beta s \lambda N}{M}\Bigl(\frac{1}{N}\sum _{i=1}^{N}\sigma_{i}^{z} \Bigr) ^{p}\Bigr]
\notag \\
&\hphantom{={}} \times
\exp \Bigl[-\frac{\beta s(1-\lambda )N}{M}\Bigl(\frac{1}{N}\sum _{i=1}^{N}\sigma_{i}^{x}\Bigr)^{2}
+
\frac{\beta (1-s)}{M} \sum _{i=1}^{N}\sigma_{i}^{x}
\Bigr] \biggr)^{M}| \{\sigma _{i}^{z}\}\rangle ,
\label{eq_s_trotter}
\end{align}
where $\sum _{\{\sigma _{i}^{z}\}}$ denotes the summation
over all spin configurations
in the $z$~basis,
and $| \{\sigma _{i}^{z}\} \rangle \equiv
\bigotimes _{i=1}^{N}| \sigma _{i}^{z}\rangle$.
The state $| \sigma _{i}^{z}\rangle$ is the eigenstate of $\sigma_{i}^{z}$,
having the eigenvalue $\sigma _{i}^{z}$ $(=\pm 1)$.
Similar notations will be used for the $x$~basis.

Following saddle point method {in the limit $N \to \infty$} and static approximation {(i.e., neglecting the imaginary-time dependence of the partition function in Eq.~(\ref{eq_s_trotter})~\cite{jorg2010energy,seki2012quantum,bapst2012quantum})},
the partition function of the system can be written as
\begin{equation}
 Z= \iint dm^{z}\,dm^{x}\, \exp [-N\beta f(\beta ,s,\lambda ;m^{z},m^{x})],
 \label{partition_function}
\end{equation}
where $f(\beta ,s,\lambda ;m^{z},m^{x})$ is the pseudo free energy defined as follows:
\begin{align}
& f(\beta ,s,\lambda ;m^{z},m^{x}) =
  (p-1)s\lambda (m^{z})^{p} -s(1-\lambda )(m^{x})^{2}
\notag \\
& -\frac{1}{\beta }\ln 2\cosh \beta
 \sqrt{\bigl\{ps\lambda (m^{z})^{p-1}\bigr\}^{2}
 + \bigl\{1-s-2s(1-\lambda )m^{x}\bigr\}^{2}}.
\label{pseudo_free energy}
\end{align}
The saddle point equations are thus
\begin{align}
& m^{z} = \frac{ps\lambda (m^{z})^{p-1}}%
  {\sqrt{
  \bigl\{ps\lambda (m^{z})^{p-1}\bigr\}^{2} + \bigl\{1-s-2s(1-\lambda )m^{x}\bigr\}^{2}
  }}
\notag \\
& \times
\tanh \beta \sqrt{
  \bigl\{ps\lambda (m^{z})^{p-1}\bigr\}^{2} + \bigl\{1-s-2s(1-\lambda )m^{x}\bigr\}^{2}},
 \label{self_consistent_eq_mz}
\\
&m^{x} = \frac{1-s-2s(1-\lambda )m^{x}}%
  {\sqrt{
  \bigl\{ps\lambda (m^{z})^{p-1}\bigr\}^{2} + \bigl\{1-s-2s(1-\lambda )m^{x}\bigr\}^{2}
  }}
\notag \\
& \times
\tanh \beta \sqrt{
  \bigl\{ps\lambda (m^{z})^{p-1}\bigr\}^{2} + \bigl\{1-s-2s(1-\lambda )m^{x}\bigr\}^{2}}.
\label{self_consistent_eq_mx}
\end{align}

To examine quantum phase transitions of the model, we consider 
low-temperature limits of the above self-consistent equations.
For a finite value of the square root in Eq.~(\ref{self_consistent_eq_mz}) and Eq.~(\ref{self_consistent_eq_mx}), the hyperbolic tangent tends 
to unity in the limit of $\beta\rightarrow\infty$. Then the 
equations are given as
\begin{align}
& m^{z} = \frac{ps\lambda (m^{z})^{p-1}}%
  {\sqrt{
  \bigl\{ps\lambda (m^{z})^{p-1}\bigr\}^{2} + \bigl\{1-s-2s(1-\lambda )m^{x}\bigr\}^{2}
  }},\label{self_consistent_eq_mz_ltemp}
\\
&m^{x} = \frac{1-s-2s(1-\lambda )m^{x}}%
  {\sqrt{
  \bigl\{ps\lambda (m^{z})^{p-1}\bigr\}^{2} + \bigl\{1-s-2s(1-\lambda )m^{x}\bigr\}^{2}
  }}.\label{self_consistent_eq_mx_ltemp}
\end{align}
In that case, the pseudo free energy~(\ref{pseudo_free energy})
becomes
\begin{align}
& f(s,\lambda ;m^{z},m^{x}) = (p-1)s\lambda (m^{z})^{p} -s(1-\lambda )(m^{x})^{2}
\notag \\
& -\sqrt{\bigl\{ps\lambda (m^{z})^{p-1}\bigr\}^{2}
 + \bigl\{1-s-2s(1-\lambda )m^{x}\bigr\}^{2}}.\label{pseudo_free energy2}
\end{align}

Equations (\ref{self_consistent_eq_mz_ltemp}) and 
(\ref{self_consistent_eq_mx_ltemp}) provide a ferromagnetic 
(${\rm F}$) solution with $m^z>0$ and a quantum paramagnetic 
(${\rm QP}$) solution for $m^z=0$ and $m^x\neq 0$. Using these 
properties of a quantum paramagnetic phase, the regions of QP 
phases can be found on the $s-\lambda$ plane. It appears 
that there exists two types of QP phases in this problem and 
we call them QP and QP2 phases to distinguish from each other. 

The regions of the different phases in terms of system parameters 
can be calculated using the above conditions of those phases 
in Eqs.~(\ref{self_consistent_eq_mz_ltemp}) and (\ref{self_consistent_eq_mx_ltemp}).
It has been found that the QP phase exists in the region $0\le s <1/(3-2\lambda )$,
and its free energy is given by
\begin{align}
 f_{\text{QP}}(s,\lambda ) = -s\lambda +2s -1,\label{f_QP_static}
\end{align}
which is independent of $p$. The free energy of the F phase can not 
be calculated analytically for a general $p$ from Eqs.~(\ref{self_consistent_eq_mz_ltemp}) 
and (\ref{self_consistent_eq_mx_ltemp}). However, in the limit 
of $p\rightarrow\infty$, the free energy of the F phase is given as
\begin{align}
f_{\text{F}}(s,\lambda )\rvert_{p\to \infty} =& -s\lambda .\label{f_F_static}
\end{align}
The free energy of the QP2 phase is given by
\begin{equation}
 f_{\text{QP2}}(s,\lambda ) = -\frac{(1-s)^{2}}{4s(1-\lambda )},\label{f_QP2_static}
\end{equation}
with the domain of applicability is restricted by $1/(3-2\lambda ) \le s < 1$.
%SSSSSSSSSSSSSSSSSSSSSSSSSSSSSSSSSSSSSSSSSSSSSSSSSSSSSSSSSSSSSSSSS

\section{Inhomogenous driving of the transverse field}
\label{inho_driving_appendix}
\subsection{Free energy and first-order transitions}
\label{free_energy_FOT}
By applying Trotter decomposition and the static approximation on 
the Hamiltonian in Eq.~(\ref{inho_driving_Ham}),
the resulting free energy at finite temperature is given 
by~\cite{susa2018exponential}
\begin{align}
&f(m; s,\tau) \nonumber\\
&~=(1-\tau)\left\{ (p-1)sm^p -T\log 2\cosh \beta\sqrt{(spm^{p-1})^2+1}\right\}
\nonumber\\
&\hspace{1cm}+\tau\left\{(p-1)sm^p -T\log 2\cosh (\beta spm^{p-1})\right\},
\label{f_Tfinite_inhomo}
\end{align}
where $m$ is the magnetization of the system along the $z$ axis. 
In the limit of zero temperature, the free energy takes form
\begin{align}
 f_0(m; s,\tau)
 &~=(1-\tau)\left\{ (p-1)sm^p-\sqrt{(spm^{p-1})^2+1}\right\} \nonumber \\
 &\hspace{1cm}+\tau\left\{(p-1)sm^p -sp m^{p-1}\right\}.
\label{f_T0_inhomo}
\end{align}
For the calculation of zero-temperature free energy it has been 
assumed that $m\ge 0$.

Using the standard Landau theory of phase transitions, the 
locations of the critical points $s_c$, $\tau_c$ 
(see Fig.~\ref{fig:phase_diagram_inhomo}) 
where the first-order transition lines terminate for 
different $p$, can be found as
\begin{equation}
    \tau_{\rm c}=\frac{1}{1+\displaystyle\sqrt{\frac{27(p-1)}{4(p-2)^3}}},\quad
    s_{\rm c}=\frac{1}{pm_{\rm c}^{p-1}\sqrt{1-{m_{1\rm c}}^2}/m_{1\rm c}},
    \label{location_first_transitions}
\end{equation}
where $m_{1\rm c}=\sqrt{(p-2)/(3(p-1))}$ and $m_{\rm c}=\tau_{\rm c}+(1-\tau_{\rm c})m_{1\rm c}$.

\subsection{Semiclassical theory of energy gap}
\label{semi_theory_EG}
One can rewrite the Hamiltonian (\ref{inho_driving_Ham}) 
in terms of two macroscopic spin operators 
(for details see Refs.~\cite{susa2018exponential,susa2018quantum}),
\begin{equation}
    S_1^{z,x}=\frac{1}{2}\sum_{i=1}^{N(1-s^r)} \sigma_i^{z,x},\quad S_2^{z,x}=\frac{1}{2}\sum_{i=N(1-s^r)+1}^{N} \sigma_i^{z,x}
\end{equation}
as
\begin{equation}
    H(s,\tau)=-sN\left\{\frac{2}{N}\big(S_1^z + S_2^z\big)\right\}^p
    -2 S_1^x.
\end{equation}
These giant operators can be considered as classical vectors for 
sufficiently large $N$, and the quantum fluctuations are 
subsequently applied around the classically stable directions 
through an expansion of the Holstein-Primakoff transformation 
to the quadratic order in terms of boson operators, as done 
in Refs.~\cite{seoane2012many,filippone2011quantum}. The result 
is given as 
\begin{align}
    H(s,\tau)=&Ne+\gamma+\frac{\delta}{2}(\sqrt{1-\epsilon^2}-1) \notag \\
    &+\Delta_{a_1} \tilde{a}_1^{\dagger}\tilde{a}_1+\Delta_{a_2} \tilde{a}_2^{\dagger}\tilde{a}_2+\Delta_{b} b^{\dagger}b,
\end{align}
where {$\tilde{a}_1$ and $\tilde{a}_2$ are bosonic annihilation operators}, and $e$ is the energy per spin of the classical ground-state. 
The parameters $\Delta_{a_1}$,  $\Delta_{a_2}$ and  $\Delta_b$ 
represent quantum fluctuations, where
\begin{align}
\Delta_{a_1} &= \delta\sqrt{1-\epsilon^2},~~~~\Delta_{a_2} = \delta.
\end{align}
Because $ \Delta_{a_2}\ge \Delta_{a_1}$, the minimum energy gap 
of the system is the smaller of $\Delta_{a_1}$ and $\Delta_{b}$: 
%\begin{align}
%    \Delta ={\rm min}(\Delta_{a_1},\Delta_{b}).
%\end{align}
%\bes
\begin{align}
    \Delta &={\rm min}(\Delta_{a_1},\Delta_{b})\nonumber \\
    \Delta_{a_1} &= \delta\sqrt{1-\epsilon^2}, ~
    \Delta_b = 2sp \{\tau+(1-\tau)\cos \theta_0 \}^{p-1},
\end{align}
%\ees
where
%\bes
\begin{align}
    \theta_0&=\arg \min_{\theta}\left\{-s[\tau+(1-\tau)\cos \theta]^p-(1-\tau)\sin \theta\right\}\nonumber\\
    \epsilon&= -\frac{2 \gamma}{\delta},\nonumber \\
    \gamma &= -\frac{1}{2}sp(p-1)(1-\tau) \sin^2 \theta_0 \{\tau+(1-\tau)\cos\theta_0 \}^{p-2},\nonumber \\
    \delta &= \Delta_b \cos\theta_0+2\sin \theta_0 +2\gamma .
\end{align}

\ethics{NA.}

\dataccess{The article has no additional data over those given in
different figures taken from published papers.}

\aucontribute{AD and BKC conceptualized the review. AR and SS
contributed to materials and organised it. All authors contributed to editing and finalizing the review.}

\competing{We declare, we do not have any competing interests.}
\funding {A.R. acknowledges UGC,
India for Start-up Research Grant No. F. 30-509/2020(BSR). The work of S.S. was supported by JSPS KAKENHI Grant No. 22K03455. A.D. acknowledges support from SPARC program, MHRD, India and SERB, DST, New Delhi, India. BKC is grateful to the Indian National Science Academy for their Senior
Scientist Research Grant.}

\ack{We acknowledge our collaborations with Gabriel Aeppli, Arunava
Chakrabarti, Arnab Das, Uma Divakaran, Jun-ichi Inoue, Sudip Mukherjee, Hidetoshi Nishimori, Masato Okada, Purusattam
Ray, Thomas F. Rosenbaum, Diptiman Sen, Parongama Sen, Robin B. Stinchcombe, Ryo Tamura and Shu Tanaka on this development.
AD acknowledges Souvik Bandyopadhyay and Sourav Bhattacharjee for comments. We are grateful to the anonymous referees for their useful comments and important suggestions.}

\disclaimer{This review is limited by our personal knowledge
and also by the size limit (which we have already  crossed). We
do not claim any  completeness  of discussions on even some
important contributions in this incredibly active field  of research.}

%%%%%%%%%% Insert bibliography here %%%%%%%%%%%%%%

\bibliographystyle{mystyle2}

\bibliography{ref_S}

\end{document}